\documentclass[aps,prd,twocolumn,noeprint,showpacs,nofootinbib,superscriptaddress,amsmath, amssymb, floatfix,table,xcdraw]{revtex4-2}
\usepackage{orcidlink}
\usepackage{graphicx}% Include figure files
\usepackage{dcolumn}% Align table columns on decimal point
\usepackage{float}
\usepackage{bm}% bold math
\usepackage{color} % Color text
\usepackage{xcolor}
\newcommand{\symfootnote}[1]{%
\let\oldthefootnote=\thefootnote%
\stepcounter{mpfootnote}%
\addtocounter{footnote}{-1}%
\renewcommand{\thefootnote}{\fnsymbol{mpfootnote}}%
\footnote{#1}%
\let\thefootnote=\oldthefootnote%
}
\newcommand{\fpeak}{$f_{\rm peak} $}

\newcommand{\fspiral}{$f_{\rm spiral} $}

\definecolor{Gray}{gray}{0.9}

\begin{document}

\preprint{}

\title{Robust and fast parameter estimation for gravitational waves\\ from binary neutron star merger remnants} 

\author{Stamatis Vretinaris
\orcidlink{0000-0001-7575-813X}} 
\email[Corresponding Author: ]{stamatis.vretinaris@ru.nl}
 \affiliation{
  Institute for Mathematics, Astrophysics and Particle Physics, Radboud University, Heyendaalseweg 135, 6525 AJ Nijmegen, The Netherlands
}

\author{Georgios Vretinaris
\orcidlink{0009-0007-6819-1586}}
\affiliation{Department of Physics, Aristotle University of Thessaloniki.} \affiliation{Mathematisches Institut, University of Tübingen, Auf der Morgenstelle 10, 72076 Tübingen, Germany}

\author{Christos Mermigkas}
 \affiliation{Department of Physics, Aristotle University of Thessaloniki.}%Lines break automatically or can be forced with \\


\author{Minas Karamanis
\orcidlink{0000-0001-9489-4612}}
 \affiliation{Berkeley Center for Cosmological Physics, University of California, Berkeley, CA 94720, USA}
 \affiliation{Lawrence Berkeley National Laboratory, 1 Cyclotron Road, Berkeley, CA 94720, USA}%Lines break automatically or can be forced with \\
 
\author{Nikolaos Stergioulas
\orcidlink{0000-0002-5490-5302}}
 \affiliation{Department of Physics, Aristotle University of Thessaloniki.}%Lines break automatically or can be forced with \\

\date{\today}% It is always \today, today,

\begin{abstract}

We present a robust and efficient methodology for parameter estimation of gravitational waves generated during the post-merger phase of binary neutron star mergers. Our approach leverages an analytic waveform model combined with empirical relations to predict prior ranges for the post-merger frequencies based on measurements of the chirp mass and effective tidal deformability in the inspiral phase. This enables robust inference of the main features of the post-merger spectrum, avoiding possible multi-modality induced by wide priors. Using waveforms derived from numerical relativity, we systematically validate our model across a broad spectrum of neutron star equations of state and mass configurations, demonstrating high fitting factors. Our method can be applied in future detections of gravitational waves from the post-merger phase with third-generation gravitational wave observatories. Furthermore, by integrating the Preconditioned Monte Carlo  sampling method within the pocoMC framework, we achieve substantial computational acceleration compared to conventional Bayesian techniques.
\end{abstract}

\maketitle

\section{Introduction} 

Gravitational Waves (GWs) have revolutionized our exploration of physical phenomena in the universe since their first direct observation in 2015 \cite{1LIGOScientific:2016aoc}. Propagating as ripples in spacetime, these waves carry invaluable information about their astrophysical sources, providing a new way to study cosmic processes. Among the various sources of GWs, Binary Neutron Star (BNS) mergers hold a special place due to their connection to multi-messenger astronomy \cite{4Sarin2021, 2022AA...665A..97R, 2022Natur.606..276H, 2024APh...15802935A, EMBNSO4} and the rich physics they reveal about the properties of neutron stars and their remnants \cite{Bauswein2016,Baiottireview,FriedmanStergioulas2020,4Sarin2021, Baiotti2022,https://doi.org/10.48550/arxiv.2104.02445}. Throughout the initial three observation runs, the LIGO-Virgo-KAGRA (LVK) collaboration \cite{LIGO,Virgo,KAGRA,KAGRA1} has documented 90 detections, predominantly involving BBH systems, with only two being BNS systems \cite{LIGOScientific2021djp}. The ongoing O4 phase is seeing an increase in the number of GW candidate events. As upgrades are made to existing detectors and third-generation detectors, such as Cosmic Explorer (CE)~\citep{CEcurves, CEscience} and Einstein Telescope (ET)~\citep{ETcurves, Branchesi2023}, are being developed, it is anticipated that a larger volume of signals will be detected, including more BNS systems \cite{KAGRA:2020npa,2021NatRP...3..344B,https://doi.org/10.48550/arxiv.2104.02445}. Here, we focus on parameter estimation of post-merger signals of BNS systems, which could provide new insights into the Equation of State (EOS) of neutron star matter.  New designs for dedicated observatories like NEMO~\cite{NEMO} and the High Frequency (HF) design proposed in~\cite{HF} are motivated by the significance of post-merger waveform signals.

The study of BNS mergers is a rapidly evolving field, with recent advancements in numerical simulations and astrophysical modeling shedding light on their various properties. The inspiral part of the GW signal provides important constraints about the EOS up to the central densities of the individual neutron stars involved in the merger (e.g. \cite{2019PhRvD.100b3012C,iacovelli2023nuclear,2024arXiv240711153C,2024PhRvD.109j3035H}), while the post-merger part will allow us to set EOS constraints closer to the maximum allowed densities, which cannot be probed by the inspiral phase, see \cite{Baiotti2017,Bauswein2019,Baiotti2019,Bernuzzi2020,3Haster2020,Diedrich2021mar,5Breschi2022,Kedia:2022nns,Bauswein:2020xlt, Breschi2024,16Criswell2023,Breschi2024,2024arXiv240303246E} and references therein.

Gravitational-wave spectra obtained from numerical-relativity simulations consistently exhibit distinct, characteristic peaks, which originate from oscillatory or transient dynamics of the surviving remnant and 
 correlate with the bulk properties of the BNS merger remnant \cite{Stergioulas2011, Hotokezaka2013, Bauswein2015, Takami2015, Bauswein2016, Bauswein2019,Paschalidis,Clark2016,Radice2016,DePietri2018,2023ApJ...952L..36F,Kastaun:2010vw}.
The most important of these peaks, can be labeled as $f_{2 - 0}$, \fspiral, \fpeak, $f_{2+0}$, in increasing order of frequency \cite{Stergioulas2011,Bauswein2015}.  The dominant peak, denoted as $f_{\rm peak}$ (or sometimes $f_2$), arises from the fundamental quadrupole oscillation excited in the remnant at the moment of merger. The peaks at frequencies $f_{2 - 0}$ and $f_{2 + 0}$ are combination tones, that arise due to the quasi-linear coupling between quadrupole and the quasi-radial oscillation modes of the remnant. The peak at the frequency $f_{\rm spiral}$  is due to transient spiral arms formed immediately after merger, that have a slower rotation rate than the core of the remnant. For further details on the various characteristics of the post-merger spectrum, see the reviews
\cite{2016EPJA...52...56B,2019AIPC.2127b0013B,2019JPhG...46k3002B,4Sarin2021,Friedman2020,annurev-nucl-013120-114541}.

 The structure of the remnant produced in a binary neutron star merger is tightly related to the bulk properties of the individual stars before merger. Thus, the post-merger frequency spectrum does not only carry the imprint of the properties of the remnant, but also of the properties of the individual stars before merger and a number of empirical relations for the dominant  post-merger frequencies have been constructed   \cite{Shibata2005,PhysRevD.71.084021, PhysRevLett.99.121102, BausweinJanka, Bauswein2012, Hotokezaka2013,Takami2014, Bauswein2015,Takami2015,Bauswein2019,FriedmanStergioulas2020,2023ApJ...952L..36F, Bernuzzi2015, bose2018, Bauswein2019, 2019PhRvD.100j4029B, VSB2020,Breschi2024}, connecting them to NS properties in the inspiral phase.

Analytical time-domain models of post-merger GW emission were presented, e.g., in \cite{Bauswein2016,bose2018,2019PhRvD.100j4029B,Easter2020,soultanis2022}.
In \cite{PhysRevD.100.043005} a hierarchical model was used to predict postmerger spectra in the frequency domain. The training set was extended in \cite{2024PhRvD.110f3008P}, where a model for the amplitude of the post-merger spectrum was constructed, using artificial neural networks.  
Frequency domain models have been developed to encompass both the inspiral and post-merger parts of the signal \cite{2022PhRvD.105j4019W,Breschi2023a,2023PhRvD.107l4009P}. Faithful models of the post-merger GW spectrum could enable more accurate parameter estimation for the post-merger part of the GW signal, revealing important information about the microphysics involved in these events \cite{Jacobi2023,2023PhRvD.108b3020V,2024arXiv240909147P,2024PhRvD.109f4032R}, including thermal effects \cite{2023ApJ...952L..36F,Raithel2023,2023ApJ...952L..36F,2024PhRvD.109d3015B,2024arXiv240102493M,2024PhRvD.110d3002R}, and the nature of phase transitions that may occur at extreme densities \cite{Fujimoto:2022xhv,Espino2024,2024PhRvD.109j3008P,2020PhRvD.102l3023B,PhysRevLett.122.061101,PhysRevLett.122.061102,Most2020,PhysRevD.104.083029}. 

The detectability of post-merger signals relies on a signal-to-noise ratio (SNR) greater than about 8, considering high-frequency noise artefacts and false alarm criteria \cite{20Panther2023}. Current detectors are not sufficiently sensitive at high frequency to allow for a direct detection of the post-merger phase \cite{Abbott:2017dke,2023CQGra..40u5008K,2024PhRvD.110h3016G}. However, future detectors are expected to allow for the first observations of post-merger signals \cite{21Zhang2023}, enabling a deeper understanding of the physics behind these events. The classification of post-merger signals is also an area of interest, as it allows for the differentiation between systems that experience prompt collapse or the formation of a neutron star remnant \cite{2Agathos2020,2023CQGra..40v5008T,Puecher2024,2024arXiv240412126J}. New numerical codes using e.g. a moving-mesh approach \cite{2024MNRAS.528.1906L} are expected to allow for more accurate determination of the post-merger dynamics.

An important aspect of extracting information from the post-merger phase is the parameter estimation of the source properties, see e.g. \cite{Easter2020}. The conventional approach to Bayesian inference typically adopts an impartial stance toward the subject being studied, relying on statistical methods like Markov Chain Monte Carlo (MCMC) algorithms to highlight prominent features \cite{Easter2020}. However, this approach encounters a limitation when there is a lack of substantial prior knowledge. 

Here, we aim to address this knowledge gap, by taking advantage of observational properties that can be extracted from the inspiral phase and using empirical relations (connecting pre- and post-merger phases) to constrain the priors in the post-merger phase, leading to a more robust parameter estimation.
In this way, we have successfully achieved accurate and robust parameter estimation, even when resampling is performed. This accomplishment has led to a significant reduction in computational time for the sampling procedure, by utilizing the Preconditioned Monte Carlo method.

The paper is organised as follows. In Section \ref{sec:methods} we review the numerical relativity data used in this work, and we present the analytic post-merger waveform model used in the parameter estimation. Furthermore we provide new empirical relations, and present our prior distributions and the preconditioned Monte carlo sampling technique. In Section \ref{sec:parameter-estimation} we present the results of our parameter estimations for various cases of signal-to-noise ratio (SNR) and for post-merger spectral types. The paper is concluded in Section \ref{sec:discussion}, which summarizes our findings.

\subsubsection{Conventions}
To refer to a particular post-merger waveform we will use the format ``EOS-Mx", where ``EOS" is the equation-of-state name and ``x" stands for the  average mass of the binary (mass of the two components when they are isolated) in solar masses.

\begin{table}
\caption{Different post-merger waveforms included in this study. The first column lists their label in the CoRe database \cite{databasepaper,CORE} (except for the last two entries). The remaining columns list the EOS name, the mass ratio and either the mass of individual components (when $q=1$) or their average mass (when $q \neq 1$).}
\begin{tabular}{|c|c|c|c|c|}
\hline
\textbf{Label}  & \textbf{EOS} & \textbf{q} & \textbf{(Average) Mass} & \textbf{References} \\ \hline
THC:0036:R03   & SLy          & 1.0        & 1.350        & \cite{Radice2016} \\ \hline
THC:0019:R05   & LS220        & 1.0        & 1.350        & \cite{Radice2017,Radice2018} \\ \hline
BAM:0088:R01   & MS1b         & 1.0        & 1.500        & \cite{databasepaper,CORE}  \\ \hline
THC:0002:R01   & BHBlp        & 1.0        & 1.300        & \cite{Radice2017,Radice2018} \\ \hline
THC:0011:R01   & DD2          & 1.0        & 1.250        & \cite{Radice2017,Radice2018} \\ \hline
BAM:0070:R01   & MS1b         & 1.0        & 1.375        & \cite{Dietrich2017} \\ \hline
BAM:0065:R03   & MS1b         & 1.0        & 1.350        & \cite{Bernuzzi2016} \\ \hline
THC:0010:R01   & DD2          & 1.0        & 1.200        &  \cite{Radice2017,Radice2018} \\ \hline
BAM:0002:R02   & 2H           & 1.0        & 1.350        & \cite{Bernuzzi2016} \\ \hline
BAM:0053:R01   & H4           & 1.5        & 1.375        & \cite{Dietrich:2016hky} \\ \hline
BAM:0124:R01   & SLy          & 1.5        & 1.250       & \cite{Dietrich2017} \\ \hline
BAM:0090:R02   & MS1b         & 1.0        & 1.600       & \cite{databasepaper,CORE}   \\ \hline
BAM:0092:R02   & MS1b         & 1.0        & 1.700    
&\cite{databasepaper,CORE}   \\ \hline
Soultanis et al.  & MPA1         & 1.0        & 1.200        & \cite{soultanis2022} \\ \hline
Soultanis et al.  & MPA1         & 1.0        & 1.550         & \cite{soultanis2022} \\ \hline
\end{tabular}
\label{table:info}

\end{table}
\begin{figure}
    \centering
    \includegraphics[width=.5\textwidth]{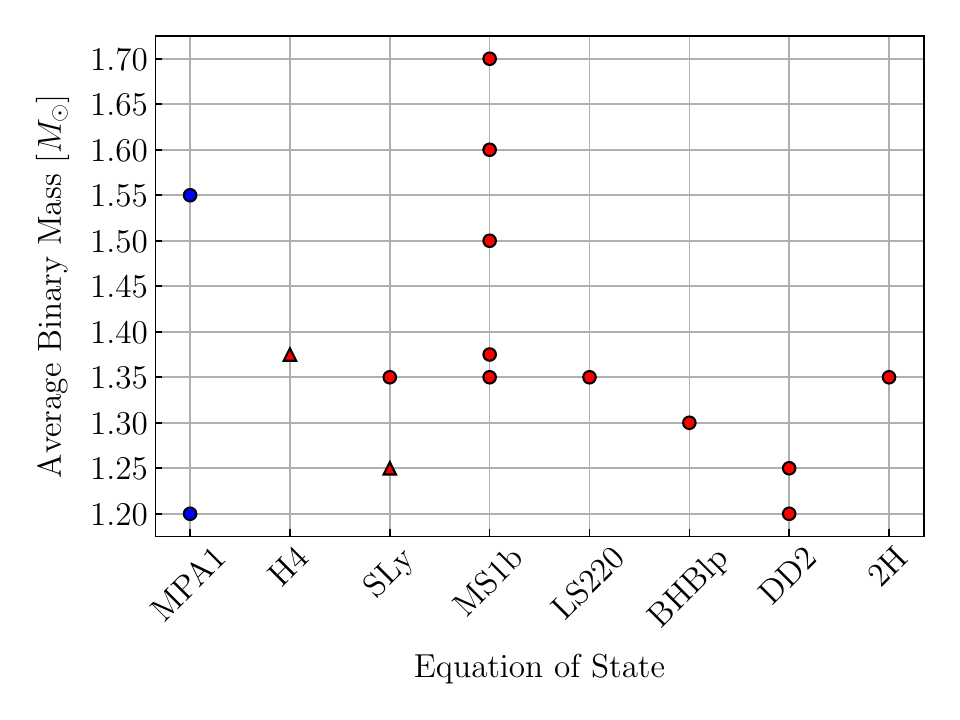}
    \caption{EOS and average binary mass configurations used in this work. Red markers correspond to waveforms from the CoRe database \cite{CORE, databasepaper} and blue markers correspond to waveforms from \cite{soultanis2022}. Circles indicate equal-mass ($q=1$) mergers, whereas triangles indicate the two cases with $q=1.5$.}
    \label{fig:waveforms}
\end{figure}

\begin{figure*}[ht]
    \centering
    \makebox[\textwidth][c]{\includegraphics[width=0.85
    \textwidth]{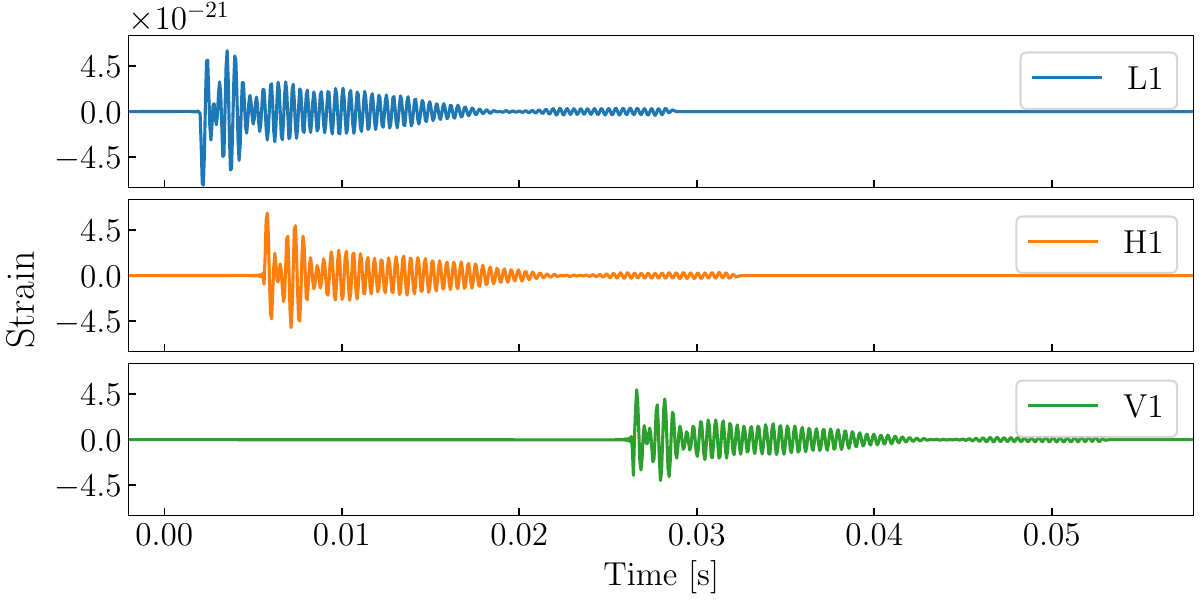}}
    \caption{An example of a projected waveform onto the L1, H1 and V1 detector network. The EOS is BHBlp and the average binary mass is 1.3$M_\odot$. The $t = 0s$ marks the time of first detection. In this case it happens for the Livingston detector.}
    \label{Fig:injections}
\end{figure*}

\section{Methods}
\label{sec:methods}

\subsection{Numerical Waveforms and Injections}
In this study we focus mainly on {\it equal mass} BNS mergers and utilize a total of 15 waveforms. Of these, 13 are waveforms from the publicly available numerical relativity catalog \texttt{CoRe} \cite{databasepaper,CORE} and the remaining two waveforms are from \cite{soultanis2022}. The waveforms are provided as $h_+(t)$ and $h_\times(t)$, the plus ($+$) and cross ($\times$) polarizations, respectively, as extracted in the numerical-relativity simulations.

Notice that our set of models is an expanded set of the 9 waveforms that were used in \cite{Easter2020} and includes those. Table \ref{table:info} lists the detailed properties of each numerical waveform, which are their label in the CoRe database (otherwise the label is Soultanis et al.), the EOS name, the mass ratio $q\geq 1$, the average mass of the two binary components and a reference to a publication where more details on the particular simulation were presented\footnote{For two of the CoRe waveforms, no particular publication exists and we cite the CoRe database.} Two of our chosen waveforms do not have equal mass, but have a relatively large mass ratio of $q=1.5$. These were included to test the robustness of our results to the mass ratio. Fig.~\ref{fig:waveforms}  shows the set of our selected numerical waves in the configuration space of average binary mass vs. EOS. Equal-mass cases are shown as circles, while the two $q=1.5$ cases are shown as triangles.

We interpolated the numerical waveforms to a uniform time step, which corresponds to a sampling frequency of 16384 Hz. To isolate the post-merger part of the signal, we truncate the waveforms at the {\it merger time}, defined as
\begin{equation}
    t_c = \text{\rm max}\{h^2_+ (t) + h^2_\times (t)  \}.
\end{equation}
To facilitate a direct comparison with the previous results of \cite{Easter2020}, we zero-pad both sides of all signals to a total duration of 0.125s, with the merger time roughly centered in the middle\footnote{P. Easter, personal communication.} 

Next, we project the post-merger parts of the numerical waveforms onto a {\it reference} three-detector network, consisting of the advanced Livingston (L1), the advanced LIGO Hanford (H1) and the advanced Virgo (V1) detectors. Fig.~\ref{Fig:injections} displays, as an example, the projected waveforms onto the L1, H1 and V1 detectors for the BHBlp EOS with $M=1.3M_\odot$. The resulting strain $h(t)$ for each detector is then {\it injected} into Gaussian noise, assuming that the detectors are operating at their design sensitivities. For the injections, we are using the \texttt{BILBY} \cite{Ashton2019} software and assume a fixed sky position\footnote{We chose the same fixed sky position and other extrinsic characteristics as in \cite{Easter2020} (P. Easter, personal communication.)}. 

We stress that we do not expect realistic detection rates for the BNS post-merger phase with such a detector network, comprising only 2G detectors at their design sensitivity. Instead, we use it only as a well-defined reference noise level. Then, we vary the distance of the source to achieve SNR values that are in the range of 8-50, which are expected to be achieved with different designs of third-generation (3G) detectors at distances of $\sim 200$Mpc \cite{2024PhRvD.110h3040B}. {\it Therefore, our results are relevant for realistic searches with 3G detector networks}\footnote{Except for the somewhat different frequency dependence of the noise curves of 3G detectors, compared to 2G detectors, in the frequency range between 1.5 and 3.5 kHz.}.

\subsection{Analytic Post-merger Waveform Model}

To perform parameter estimation for post-merger injections, we will use an analytic waveform model, which is an extension of the model presented in \cite{Easter2020} and includes four damped oscillators and a linear drift for each frequency. Our analytic post-merger waveform model reads 
\begin{align}
    h(\boldsymbol{\theta}, t) & = h_+(\boldsymbol{\theta}, t) - i h_\times (\boldsymbol{\theta},t) \nonumber \\
                 & = \sum_{j=1}^4 \left[ h_{j,+}(\boldsymbol{\theta},t) - i h_{j,\times} (\boldsymbol{\theta},t) \right],
                 \label{Eq:anmodel}
\end{align}
with the individual ($+$) polarization components written as
\begin{equation}
    h_{j,+}(\boldsymbol{\theta}, t) = A_j \exp\left[ -\frac{t}{T_j}\right] \cos\left[2\pi f_j t(1+a_j t) + \psi_j\right].
    \label{Eq:analytic}
\end{equation}

In Eq. \ref{Eq:analytic}, $\boldsymbol{\theta}$ is a vector containing the intrinsic parameters of the model
\[
\boldsymbol{\theta} = \left\{ A_j, T_j, f_j, a_j,\psi_j\right\} \quad \text{for}\quad  j\in[1,4],
\]
where $A_j$ are the individual amplitudes, $f_j$ the frequencies, $T_j$ the damping times, $\psi_j$ the phases, and $a_j$ the frequency drift coefficients. 
The ($\times$) polarization components are then produced by applying a $\pi/2$ phase shift on the respective ($+$) polarization components.

Compared to the analytic model used in \cite{Easter2020}, our model has two main differences:
\begin{itemize}
    \item we have added a fourth oscillator, $f_4$ at frequencies $>f_{\rm peak}$, which could correspond to e.g. $f_{2+0}$ \cite{Stergioulas2011} when this combination tone is present with sufficient amplitude, or to some other high-frequency peak in the post-merger spectrum, and
    \item we are using the amplitudes $A_j$ as free parameters, whereas in \cite{Easter2020} an additional constraint of $A_1+A_2+A_3=H$ was imposed, where $H$ was some global amplitude. 
\end{itemize}
In our investigation, we found that adopting the $A_1+A_2+A_3=H$ constraint imposes a \textit{simplex} geometry and may introduce {\it degeneracies} \cite{Betancourt2012} in the posterior distribution of parameter estimation. 

{\begin{figure}[ht]
    \centering
    \includegraphics[width=.49\textwidth]{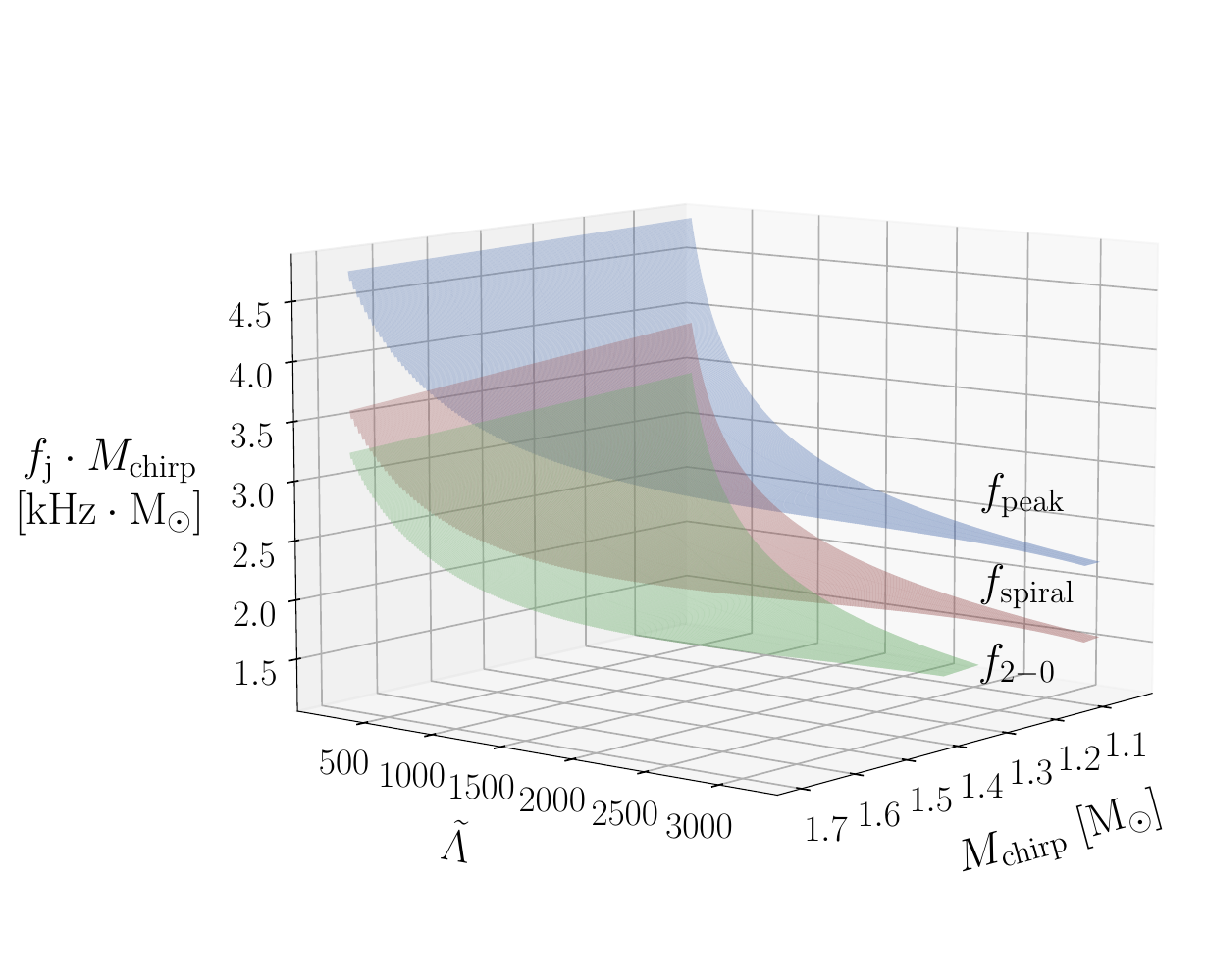}
    \caption{Empirical surfaces for the three most important post-merger frequencies vs. $M_{\mathrm{chirp}}$ and $\tilde{\Lambda}$. The blue surface corresponds to $f_{\mathrm{peak}}$, the red to $f_{\mathrm{spiral}}$ and the green to $f_{2-0}$.}
    \label{fig:empirical-surfaces}
\end{figure}}

\subsection{Empirical Relations and Classification}
\label{section:ER}

Here, we use empirical relations for some of the post-merger frequencies to tighten the prior distribution in the Bayesian parameter estimation. As we show in Sec. \ref{sec:parameter-estimation}, this leads to a robust parameter estimation, overcoming the problem of multi-peak distributions obtained when a flat prior distribution over a wide frequency range is assumed for the post-merger peaks (see, e.g., Figs. C1, C5 in \cite{Easter2020}).

\begin{figure}[h!]
    \centering
    \includegraphics[width=.48\textwidth]{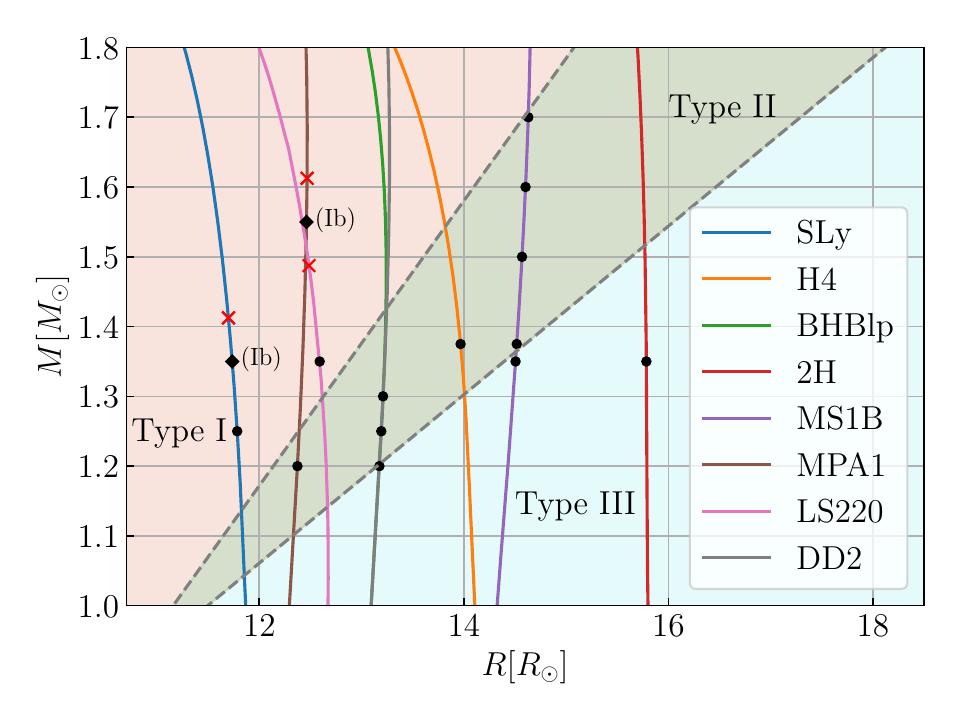}
    \caption{Mass-Radius diagram of equilibrium sequences constructed with the different EOS in Table \ref{table:info}. For each model BNS system used in this study, the average mass of its two components is shown with black markers. We use the classification scheme of \cite{Bauswein2015,VSB2020} to indicate models that give rise to post-merger spectra of Type I (light red region), Type II (light green region) and Type III (light blue region), see text for details. Red crosses denote 1/2 of the threshold mass to prompt collapse. Models close to the threshold mass (shown here as Type Ib) exhibit a different behavior, see \cite{soultanis2022}.}
    \label{fig:classification}
\end{figure}

For the primary post-merger frequency, a useful empirical relation for  $f_{\mathrm{peak}}$ was derived in \cite{VSB2020}:
\begin{equation}\label{eq:peak}
f_\mathrm{peak}M_\mathrm{chirp} =  1.392 - 0.108 M_\mathrm{chirp}+51.70 \tilde{\Lambda}^{-1/2},
\end{equation}
where $M_\mathrm{chirp} = (M_A M_B)^{3 / 5}/(M_A+M_B)^{1 / 5}$, is the chirp mass of the binary, with $M_A, M_B$ being the individual masses and 
\begin{equation}
\tilde{\Lambda}=\frac{16}{13} \frac{\left(M_A+12 M_B\right) M_A^4 \Lambda_A+\left(M_B+12 M_A\right) M_B^4 \Lambda_B}{\left(M_A+M_B\right)^5}
\end{equation}
is the mass-weighted tidal deformability, with $\Lambda_A, \Lambda_B$ being the individual dimensionless tidal deformabilities, respectively. This empirical relation has a maximum residual of $0.302$kHz and $R^2=0.985$. It was constructed using a large set of both equal-mass and un-equal-mass models and is thus relatively insensitive to the mass ratio $q$. 

We use the particular empirical relation in Eq. (\ref{eq:peak}), because both $M_\mathrm{chirp}$ and $\tilde{\Lambda}$ can be constrained by GW observations in the inspiral phase of a BNS merger. Using the data in Table II of \cite{VSB2020} and a standard Levenberg-Marquardt nonlinear least-squares minimization algorithm, we construct similar empirical relations for two additional post-merger peaks:
\begin{equation}\label{eq:quasiradial}
    f_\mathrm{2-0}M_\mathrm{chirp} =  0.558 - 0.406 M_\mathrm{chirp}+48.696 \tilde{\Lambda}^{-1/2},
\end{equation}
with a maximum residual of $0.362$kHz and $R^2=0.941$, and
\begin{equation}\label{eq:spiral}
f_\mathrm{spiral}M_\mathrm{chirp} =  1.2 - 0.442 M_\mathrm{chirp}+45.819 \tilde{\Lambda}^{-1/2},
\end{equation}
with a maximum residual of $0.461$kHz and $R^2=0.944$.
Fig.~\ref{fig:empirical-surfaces} displays all three empirical relations in Eqs. (\ref{eq:peak}), (\ref{eq:spiral}) and (\ref{eq:quasiradial}) as two-dimensional, non-intersecting surfaces.

Post-merger spectra can be classified into three main types (I, II, and III), based on the relative strength of the secondary frequencies $f_{2-0}$ and $f_{\rm spiral}$ ~\cite{Bauswein2015,VSB2020}.
Fig.~\ref{fig:classification} shows the mass-radius diagram of equilibrium sequences constructed with the different EOS in Table \ref{table:info}. For each model BNS system used in this study, the average mass of its two components is shown with black markers. The region of models with Type I post-merger spectra is shown in light red, with Type II spectra in light green, and with Type III spectra in light blue color. 

Notice that the boundaries between the regions of different post-merger types in Fig.~\ref{fig:classification} are not sharp. Rather, there is a continuous transition between the different types. In addition, in \cite{soultanis2022} a different behavior was noticed for merger remnants that come close to the threshold mass for prompt collapse\footnote{The values of the threshold mass for prompt collapse can be found in Table 1 of \cite{Bauswein2020}.} (see the red crosses in Fig.~\ref{fig:classification}). For these models, both $f_{2-0}$ and $f_{\rm spiral}$ are important, and there is partial overlap of the peaks in the frequency spectrum. This is because the quasi-radial frequency decreases rapidly, as the quasi-radial instability is approached; see \cite{2013rrs..book.....F}. {\emph We indicate such models as Type Ib} in Fig.~\ref{fig:classification}.

Depending on the particular BNS merger, fewer than four damped oscillators may be meaningful for inclusion in the analytic post-merger model of Eq. (\ref{Eq:anmodel}). In our Bayesian parameter estimation, we use the classification scheme of \cite{Bauswein2015,VSB2020}, as applied in Fig.~\ref{fig:classification}, to decide which secondary frequencies will be constrained through their respective empirical relations and which will be assumed to have a uniform prior in a wide frequency range (see Sec. \ref{Sec:priors}).

\subsection{Bayesian inference}

Our Bayesian parameter estimation is based on the likelihood function
\begin{equation}
\mathcal{L}(d \mid \boldsymbol{\theta}) \propto \exp \bigl[-\bigl< d(t)-h(\boldsymbol{\theta}, t), d(t)-h(\boldsymbol{\theta}, t)\bigr> \bigr],
    \label{Eq:L}
\end{equation}
where $d(t) = h(t) +n(t)$ is the {\it injected waveform} in each detector, that is, the sum of the projected numerical waveform $h(t)$ and the noise $n(t)$ into which it is injected. In Eq. (\ref{Eq:L}), the {\it noise-weighted inner product} is 
\begin{equation}
   \bigl< h, g \bigr> = 4 \mathfrak{Re} \int_0^\infty \frac{\tilde{h}(f) \tilde{g}^\ast(f)}{S(f)}df,
\end{equation}
where a tilde denotes the Fourier transform, an asterisk denotes complex conjugations and $S(f)$ is the detector's noise power spectral density. 

To evaluate the effectiveness of our parameter estimation, we compute the {\it noise-weighted fitting factor} $\mathcal{F}$ \cite{apostolatos95} between an injected waveform $d(t)$ and a {\it posterior} waveform $h(\boldsymbol{\theta},t)$
\begin{equation}
	\mathcal{F}(d(t),h(\boldsymbol{\theta},t))\equiv\frac{
\bigl<d(t)|h(\boldsymbol{\theta},t)\bigr>}{\sqrt{\bigl<d(t)|d(t)\bigr>\bigl<h(\boldsymbol{\theta},t)|h(\boldsymbol{\theta},t)\bigr>}}.
	\label{eq:FF}
\end{equation}
For the projected numerical waveforms $h(t)$,
we use \texttt{pyCBC} \cite{alexnitz20237885796} to calculate their optimal SNR for a single detector as
\begin{equation}
\rho_{{\rm opt},i}=\bigl< h, h\bigr>^{1/2},
\end{equation}
and the network SNR as
\begin{equation}
\rho_{\rm opt}= \left(\sum_{i \in \mathrm{HLV}} \rho_{{\rm opt},i}^2 \right)^{1/2}.
\end{equation}
As already mentioned in Sect. \ref{sec:methods}, we scale the source distance to obtain a desired network SNR in the range of 8 to 50, corresponding to the expected event rates for a network of 3G detectors.

\subsection{Informative Gaussian Priors}
\label{Sec:priors}

Gravitational waves from the post-merger phase of individual BNS mergers are expected to be detected for systems that are loud enough to have a post-merger network SNR larger than about 8. This means that the inspiral part of the network will have a significantly larger SNR, allowing for a very accurate determination of the chirp mass $M_{\rm chirp}$, as well as fairly good constraints on the mass-weighted tidal deformability $\tilde \Lambda$ (see \cite{iacovelli2023nuclear,2024PhRvD.109j3035H} for recent estimates).
Here, as a proof of principle, we will assume that the binary system's $M_{\rm chirp}$ and $\tilde \Lambda$ have been obtained from the inspiral part of the waveform with accuracy better than the residuals of the empirical relations in Eqs. (\ref{eq:peak}), (\ref{eq:quasiradial}) and (\ref{eq:spiral}). In future work, realistic distributions for the measured $M_{\rm chirp}$ and $\tilde \Lambda$, depending on the expected SNR for next-generation detectors, should be incorporated in this analysis.

Given $M_{\rm chirp}$ and $\tilde \Lambda$ from the inspiral phase, we can calculate the expected frequencies for $f_{\rm peak}$, $f_{\rm 2-0}$ and $f_{\rm spiral}$ from the empirical relations in Eqs. (\ref{eq:peak}), (\ref{eq:quasiradial}) and (\ref{eq:spiral}), respectively. These can then be used to set informative priors for some of the frequencies of our analytic model in the Bayesian parameter estimation. Specifically, we assume a Gaussian normal distribution with mean value equal to the prediction of the respective empirical relation and a standard deviation ($\sigma$), chosen such that the maximum residual in the empirical relations in Eqs. (\ref{eq:peak}), (\ref{eq:quasiradial}) and (\ref{eq:spiral}) is equal to $3\sigma$.

We always assume a Gaussian normal prior distribution for the $f_{\rm peak}$ frequency and a uniform prior in the range $(f_{\rm min}, 5{\rm kHz})$, where $f_{\rm min}=f_{\rm peak}+0.3{\rm kHz}$ for the frequency $f_4$ of the fourth damped oscillator (with frequency $>f_{\mathrm{peak}}$) in our analytic model (\ref{Eq:anmodel}). For the other two frequencies, we use the classification scheme discussed in Sec. \ref{section:ER} to set priors, according to the type of post-merger spectrum, as follows:
\begin{itemize}
    \item Type I: Gaussian priors, ${\cal N}(f_{2-0},\sigma^2)$ for $f_{2-0}$ and uniform priors ${\cal U}(1, 5)$[kHz] for $f_{\rm spiral}$. 
     \item Type II: Gaussian priors, ${\cal N}(f_{2-0},\sigma^2)$ 
     for $f_{\rm 2-0}$ and ${\cal N}(f_{\rm spiral},\sigma^2)$ for $f_{\rm spiral}$.
     \item Type III: Gaussian priors, ${\cal N}(f_{\rm spiral},\sigma^2)$ for $f_{\rm spiral}$ and uniform priors ${\cal U}(1, 5)$[kHz] for $f_{\rm 2-0}$. 
\end{itemize}
Notice that when we use uniform priors for $f_{2-0}$ or $f_{\rm spiral}$, we adopt the same frequency range as in \cite{Easter2020}. The reason we adopt uniform priors in the range 1-5kHz for $f_{2-0}$ or $f_{\rm spiral}$, when these peaks are expected to be very weak (based on the post-merge type), is not that we expect them to have frequencies outside the range predicted by the corresponding empirical relation, but to give the sampler the opportunity to explore other features of the spectrum, where it may be more meaningful to fit a damped sinusoid than to use it to fit a very weak $f_{2-0}$ or $f_{\rm spiral}$.

The priors for the amplitudes and frequency drift terms in the analytic model are set to $A_j \sim
\mathcal{U}(-24,-19)$ and to $\alpha_j \sim \mathcal{U}(-6.4,6.4)$, respectively. The priors for the remaining intrinsic parameters in $\boldsymbol{\theta}$ are set as in \cite{Easter2020}.

\subsection{Preconditioned Monte Carlo Sampling}
\label{section:poco}

As the post-merger signals that we fit are relatively weak and the model nonlinear, we expect the posterior distribution of the model parameters to be non--Gaussian and potentially multimodal in specific cases. 
These deviations from Gaussianity can impede the effectiveness of most sampling techniques,including many Markov chain Monte Carlo (MCMC) and nested sampling variants, to produce independent samples from the posterior distribution. 
To overcome this challenge, we used the recently developed method of Preconditioned Monte Carlo (PMC) as implemented in the open source \texttt{ pocoMC} Python package \cite{karamanis2022accelerating, karamanis2022pocomc}. 
{\it PMC utilizes a normalizing flow to iteratively decorrelate the parameters of the posterior distribution},  effectively Gaussianizing it. 

PMC targets an annealed version of the posterior, with density given by
\begin{equation}
    \label{eq:annealed_posterior}
    p_{t}(\boldsymbol{\theta}\vert d) \propto \mathcal{L}^{\beta_{t}}(\boldsymbol{\theta}\vert d) \pi(\boldsymbol{\theta}),
\end{equation}
where $\beta_{t}$ is the parameter controlling the annealing process.
A collection of particles gradually transitions from the prior (i.e. $\beta_{0}=0$) to the posterior distribution (i.e. $\beta_{T}=1$) through a sequence of reweighting, resampling, and mutation steps. 
After each iteration, the normalizing flow is fitted to the distribution of the particles and the sampling proceeds in the uncorrelated latent space of the normalizing flow. 
Since the transformed annealed posterior distribution exhibits almost no correlations in the latent space, the sampling performance is substantially increased. 
The sampling procedure ends when the annealing parameter $\beta$ reaches the value of unity, which corresponds to the actual posterior distribution.

In our application, \texttt{pocoMC} generates the required number of samples almost {\it an order of magnitude faster} in terms of the total number of likelihood evaluations and the wall clock time, compared to nested sampling using \texttt{dynesty} \cite{Speagle2020}.

\begin{figure}[t]
    \centering
    \includegraphics[width=.49\textwidth]{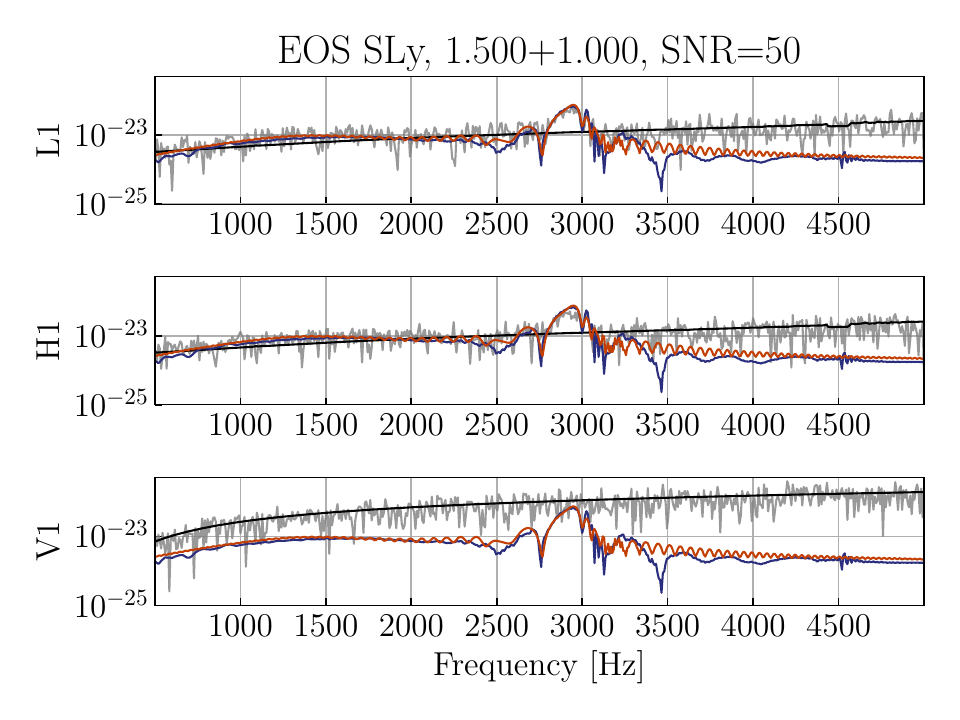}
    \caption{Amplitude spectral density (ASD) in $1/\sqrt{\rm Hz}$ of the injected waveforms (blue lines) in Gaussian noise around the ASD of each of the three detectors comprising the network. Grey curves represent the sum of the noise plus the injected waveforms. Red curves are the maximum likelihood reconstructed waveforms.}
       \label{fig:SLy_1.5_1_ASD}
\end{figure}

\begin{figure*}[ht]%
    \centering
    \includegraphics[width=.45\textwidth]{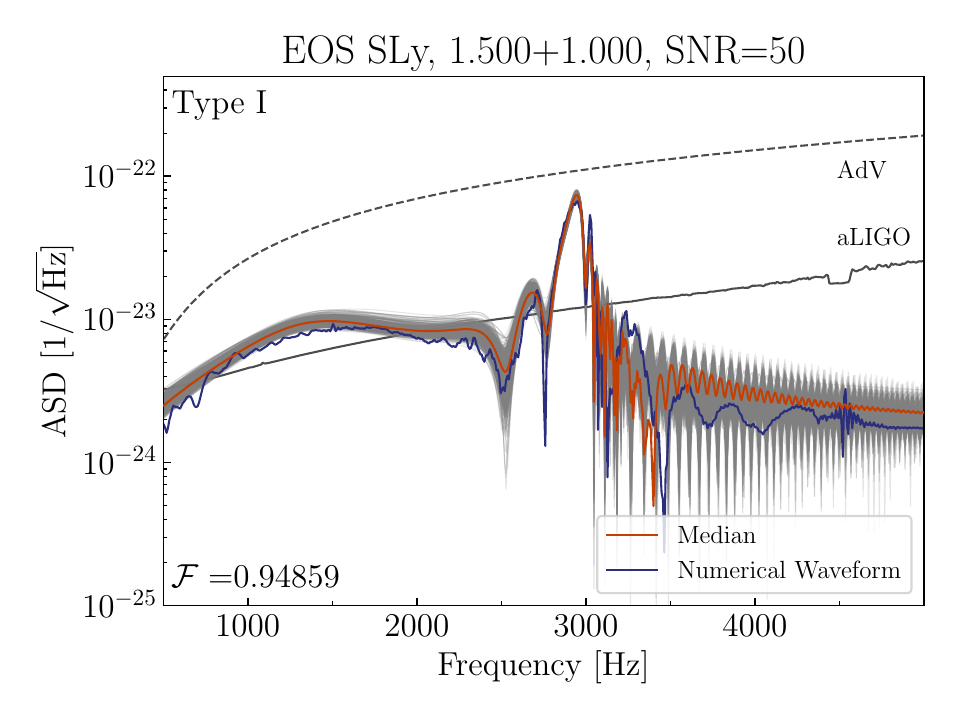}
    \qquad
    \includegraphics[width=.50\textwidth]{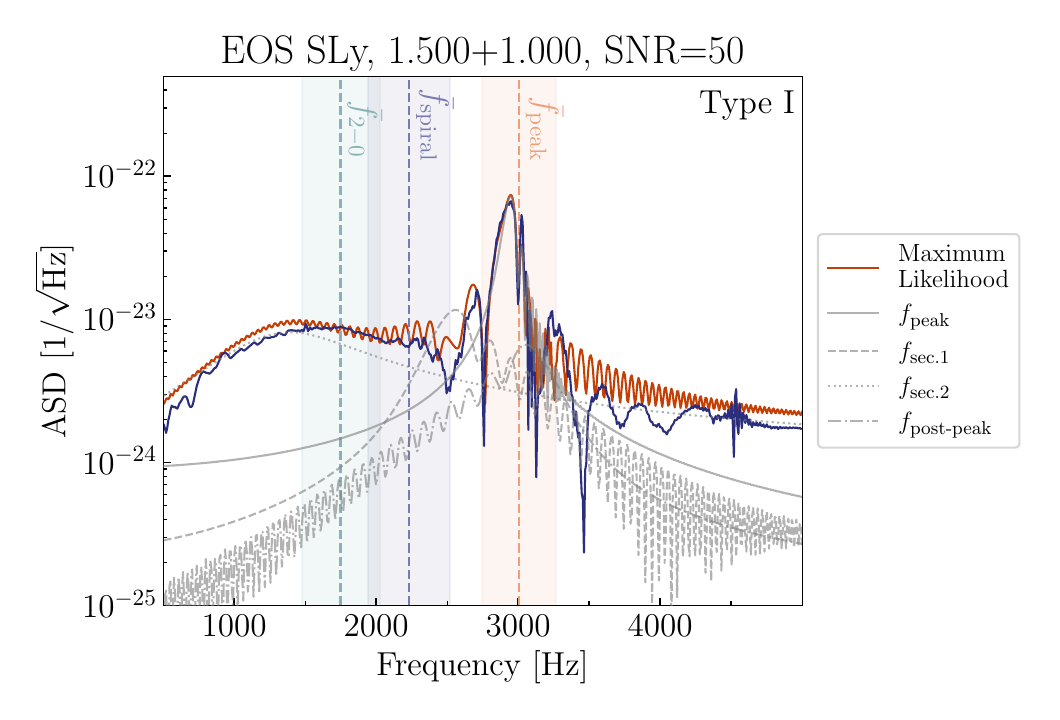}
    \caption{Amplitude spectral density (ASD) of gravitational wave signals from a binary neutron star merger with EOS MPA1 and equal component masses of 1.550 solar masses. The left panel compares the ASD to the sensitivity of aLIGO and AdV detectors, while the right panel analyzes the spectral features of the signal with respect to various characteristic frequencies, at an SNR of 50. The fitting factor $\cal F$ is noted, suggesting the closeness of the numerical waveform to the median signal.}%
    \label{fig:SLy_1.5_1_medmax}%
\end{figure*}

\section{PARAMETER ESTIMATION}
\label{sec:parameter-estimation}
The cases studied in this work have a network SNR of 50, 16 and 8. This is motivated by 3G detector designs that can reach post-merger network SNR's up to O(50) at distances of 200Mpc \cite{CEcurves}.
%{\color{blue} add more citations and comment on the event rate at such a distance.}
We start with the case of ${\rm SNR}=50$ and  discuss the parameter estimation obtained for four representative cases, one for each of the four types I, Ib, II and III, according to the classification discussed in Sec. \ref{section:ER}.

\subsection{${\rm SNR}=50$} \label{SNR50}

\subsubsection{Type I (EOS SLy, 1.5+1.0 $M_\odot$)} \label{sec:typeI}

As a first example, we discuss the case of a 1.5+1.0 $M_\odot$ merger with EOS SLy. This EOS is very soft and the model lies well in the Type I region of Fig.~\ref{fig:classification}, i.e. $f_{\rm 2-0}$ is the dominant secondary peak in the post-merger spectrum\footnote{Because the mass ratio is $q=1.5$, the amplitude of the $f_{\rm spiral}$ contribution is even weaker than it would have been for an equal mass case, since the $f_{\rm spiral}$ contribution is created by antipodal tidal bulges that are strong emitters when they are symmetric \cite{Bauswein2015}.}. Consequently, our scheme leads to Gaussian priors ${\cal N}(f_{\rm peak},\sigma^2)$, ${\cal N}(f_{\rm 2-0},\sigma^2)$ centered on the predictions of the empirical relations (\ref{eq:peak}) and (\ref{eq:quasiradial}) for $f_{\rm peak}$ and $f_{\rm 2-0}$, correspondingly, and uniform priors ${\cal U}(1, 5)$[kHz] for $f_{\rm spiral}$. The fourth oscillator, $f_4$, in our analytic model (\ref{Eq:anmodel}) is initialized with uniform priors in the range $(f_{\rm min}, 5{\rm kHz})$, where $f_{\rm min}=f_{\rm peak}+0.3{\rm kHz}$.

In Fig.~\ref{fig:SLy_1.5_1_ASD}, we show the amplitude spectral density (ASD) of the injected waveforms (blue lines) in Gaussian noise around the ASD of each of the three detectors comprising the network \cite{LIGO2010,LIGO2016}. The gray curves in this plot represent the sum of the noise plus the injected waveforms. The maximum likelihood reconstructed waveforms (red curves) agree to a high degree with the injected waveforms.

In the left panel of Fig.~\ref{fig:SLy_1.5_1_medmax}, we display the ASD of 1000 randomly sampled waveforms (gray lines), constructed using the posterior distribution of the inferred parameters of our analytic model of Eq. (\ref{Eq:anmodel}). The ASDs of the random reconstructed waveforms cluster around their median (red line), in good agreement with the ASD of the numerical waveform (blue line). Some oscillations can be seen at high frequencies, which have only a small impact on the overall agreement between the reconstruction and the numerical waveform. The fitting factor between the median reconstructed waveform and the numerical waveform is ${\cal F} = 0.94859$.

The right panel of Fig.~\ref{fig:SLy_1.5_1_medmax} shows, in more detail, the individual contributions of the different damped oscillators included in the analytic model of Eq. (\ref{Eq:anmodel}). Here, the red line represents the maximum likelihood reconstructed waveform, whereas the different gray lines correspond to the individual damped oscillators. Vertical dashed lines represent the predicted frequencies using the empirical relations (\ref{eq:peak}), (\ref{eq:quasiradial}) and (\ref{eq:spiral}), for $f_{\rm peak}$, $f_{\rm 2-0}$ and $f_{\rm spiral}$, respectively. The colored bands indicate the maximum error for each of the three empirical relations.

 \begin{figure*}%
    \centering
    \includegraphics[width=.45\textwidth]{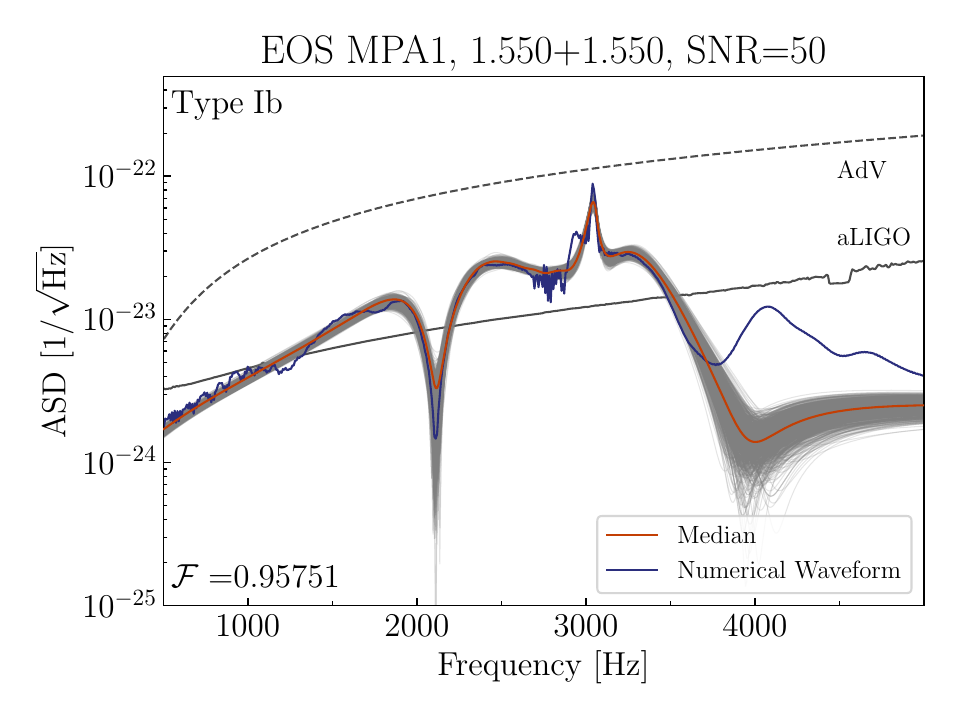} %
    \qquad
    \includegraphics[width=.50\textwidth]{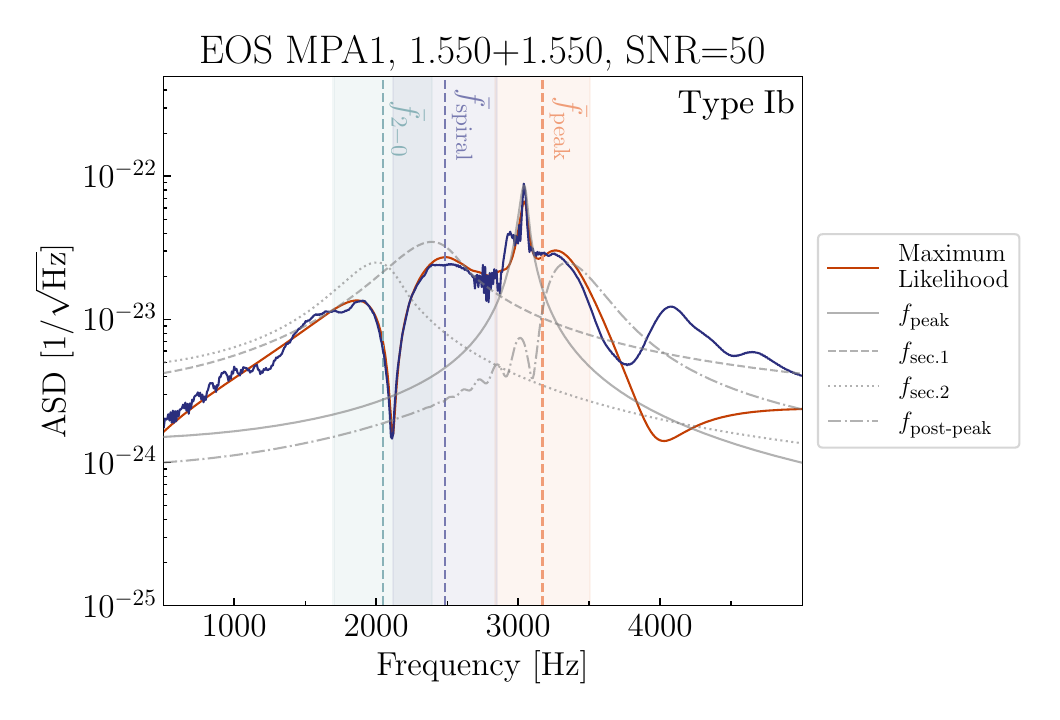} %
    \caption{Same as Fig. \ref{fig:SLy_1.5_1_medmax}, but for a 1.55+1.55 $M_\odot$ model with EOS MPA1 (Type Ib post-merger spectrum).}%
    \label{fig:/MPA1-1.55_medmax}%
\end{figure*}

We observe in the right panel of Fig.~\ref{fig:SLy_1.5_1_medmax} that the damped oscillator corresponding to $f_{\rm peak}$ describes the shape of $f_{\rm peak}$ with high accuracy. On the other hand, $f_{\rm spiral}$ can barely be seen in this spectrum and the damped oscillator corresponding to $f_{\rm spiral}$ (shown as $f_{\rm sec.1}$ in the figure), has converged to a frequency at the upper edge of the frequency band of the $f_{\rm spiral}$ empirical relation, to accommodate another peak in between\footnote{Notice that the classification scheme of \cite{Bauswein2015} has not yet been fully extended to the case of large mass ratios and such additional features still need to be fully explained and taken into account in analytic models.} $f_{\rm spiral}$ and $f_{\rm peak}$. The $f_{\rm 2-0}$ appears prominently in this Type I spectrum, and the corresponding damped oscillator (shown as $f_{\rm sec.2}$ in the figure) converges at the left edge of the corresponding empirical frequency band.

 At frequencies higher than $f_{\rm peak}$, the strongest peak is very close to $f_{\rm peak}$. In fact, the two features at frequencies somewhat smaller and somewhat higher than $f_{\rm peak}$ seem to form a triplet with the latter. This could be due to modulations induced in $f_{\rm peak}$ by the evolution of the remnant created by the unequal mass merger. The fourth oscillator (shown as $f_{\rm post-peak}$ in the figure) converges to a frequency in between the main $f_{\rm peak}$ and the feature at somewhat higher frequency. 

 Overall, for this Type I model, the $f_{\rm peak}$ and $f_{\rm 2-0}$ frequency peaks are well described by the corresponding damped oscillators, whereas the other two oscillators in the analytic model converge close to other prominent features, in a way that the sum of all four (shown in red as the maximum likelihood waveform) achieves a high fitting factor.

\subsubsection{Type Ib (EOS MPA1, 1.55+1.55 $M_\odot$)} \label{sec:typeIb}

\begin{figure*}%
    \centering
    \includegraphics[width=.45\textwidth]{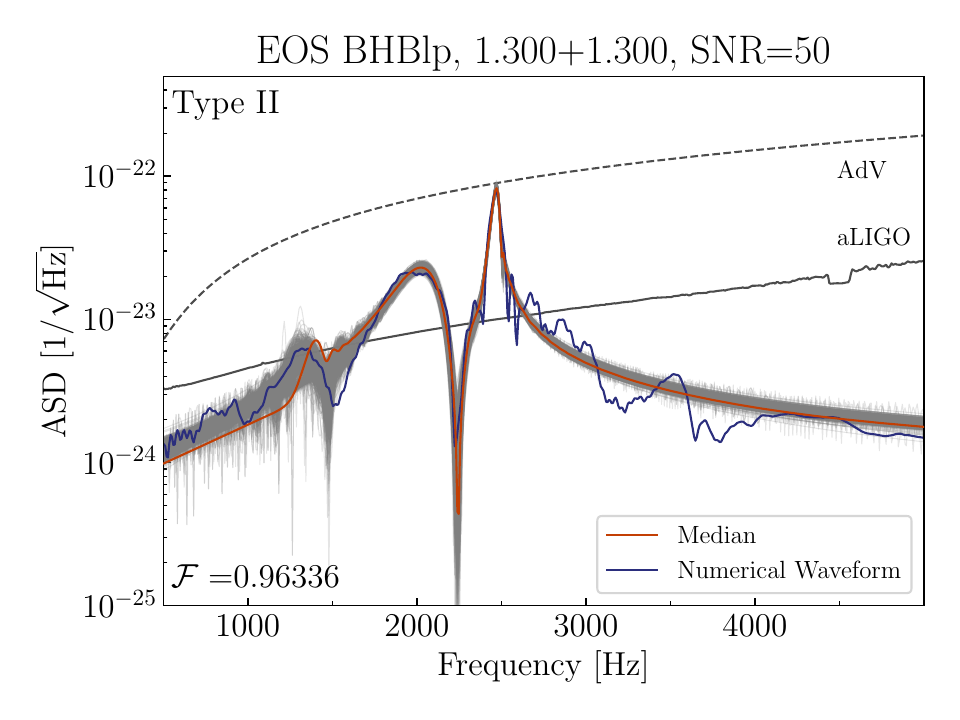}
    \qquad
    \includegraphics[width=.50\textwidth]{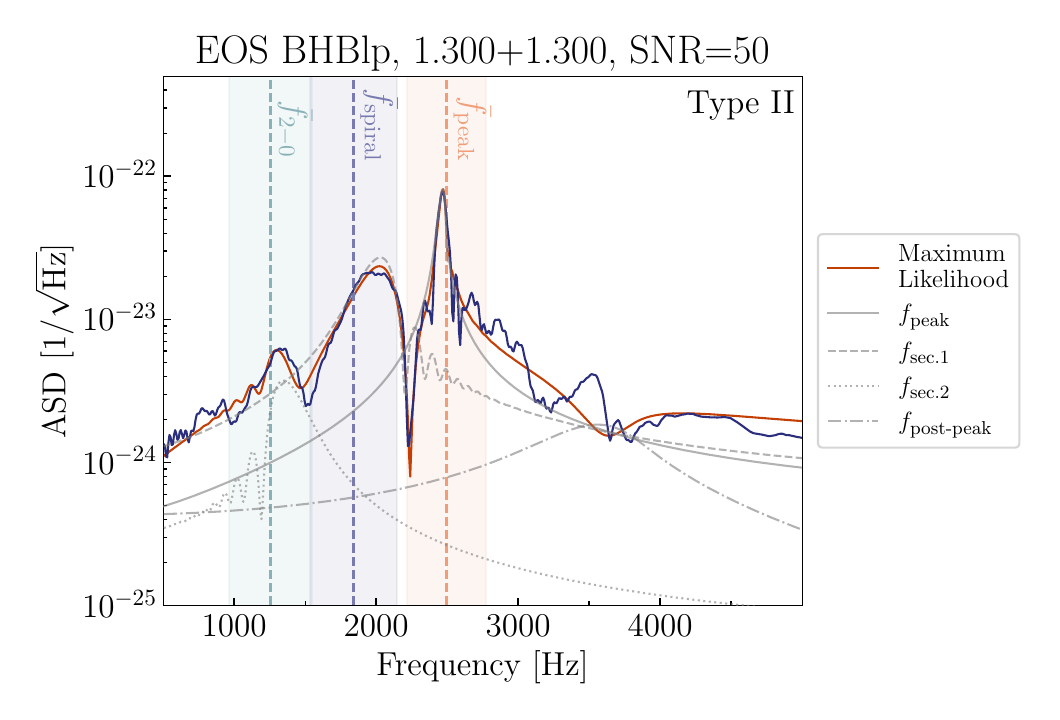}
    \caption{Same as Fig. \ref{fig:SLy_1.5_1_medmax}, but for a 1.3+1.3 $M_\odot$ model with EOS BHBlp (Type II post-merger spectrum).}%
    \label{fig:BHBlp-1.3_medmax}%
\end{figure*}

Next, we discuss a representative case of a Type Ib spectrum. This is a 1.55+1.55$M_\odot$ merger using the MPA1 EOS. Its average mass is close to the threshold mass for prompt collapse to black holes (see Fig.~\ref{fig:BHBlp-1.3_medmax}). It was already discussed in \cite{soultanis2022} that in such cases, the $f_{2-0}$ peak appears closer to $f_{\rm peak}$ and practically merges with the $f_{\rm spiral}$ frequency, because the quasi-radial frequency $f_0$ becomes smaller, as the threshold to prompt black hole formation is approached. These are outlier cases, which are at the edge of the maximum error of the empirical relation for $f_{2-0}$. A second prominent feature of this particular model is the appearance of a late-time ($t-t_{\rm merger}>10{\rm ms}$) rotational instability; see Fig.~27 in \cite{soultanis2022} and the detailed analysis in \cite{Sasli2023wpa}. In the post-merger GW spectrum, this adds a characteristic narrow peak to the broader $f_{\rm peak}$ of the first 10-15ms after merger.

In the left panel of Fig.~\ref{fig:/MPA1-1.55_medmax}, the 1000 randomly sampled waveforms (gray lines), constructed using the posterior distribution of the inferred parameters of our analytic model of Eq. (\ref{Eq:anmodel}) cluster tightly around their median (red line), in good agreement with the ASD of the numerical waveform (blue line), except for very high frequencies, where the detectors are not sentitive. The fitting factor between the median reconstructed waveform and the numerical waveform is ${\cal F} = 0.95751$.

In the right panel of Fig.~\ref{fig:/MPA1-1.55_medmax}, we observe that the damped oscillator corresponding to $f_{\rm peak}$ has converged specifically to the narrow $t-t_{\rm merger}>10{\rm ms}$ contribution of the rotational instability. The broader $f_{\rm peak}$ contribution of the first 10ms after the merger is described mainly by the fourth oscillator, shown as $f_{\rm post-peak}$ in this figure. On the other hand, the first secondary oscillator ($f_{\rm sec.1}$) describes relatively well the merged $f_{2-0}$ and $f_{\rm spiral}$ frequencies\footnote{We stress that the $f_{2-0}$ peak is not at the center of the empirical band for $f_{2-0}$ in Fig.~\ref{fig:/MPA1-1.55_medmax}, but at the right edge of this band and has merged with $f_{\rm spiral}$.}. Finally, the other secondary oscillator ($f_{\rm sec.2}$) has converged to a low-frequency peak that is not part of the classification of \cite{Bauswein2015} (for soft EOSs, $f_{\rm peak}$ can be so high, that additional features of the post-merger GW spectrum can appear at frequencies higher than the highest frequency of the inspiral phase).

Overall, we see that the analytic model of Eq. (\ref{Eq:anmodel}), combined with our scheme for choosing the Gaussian priors based on the empirical relations in Eqs. (\ref{eq:peak}), (\ref{eq:quasiradial}) and (\ref{eq:spiral}), and the classification scheme of \cite{Bauswein2015} has sufficient flexibility to capture all essential features of the post-merger GW spectrum of Type Ib models, achieving a high fitting factor.

\begin{figure*}%
    \centering
    \includegraphics[width=.45\textwidth]{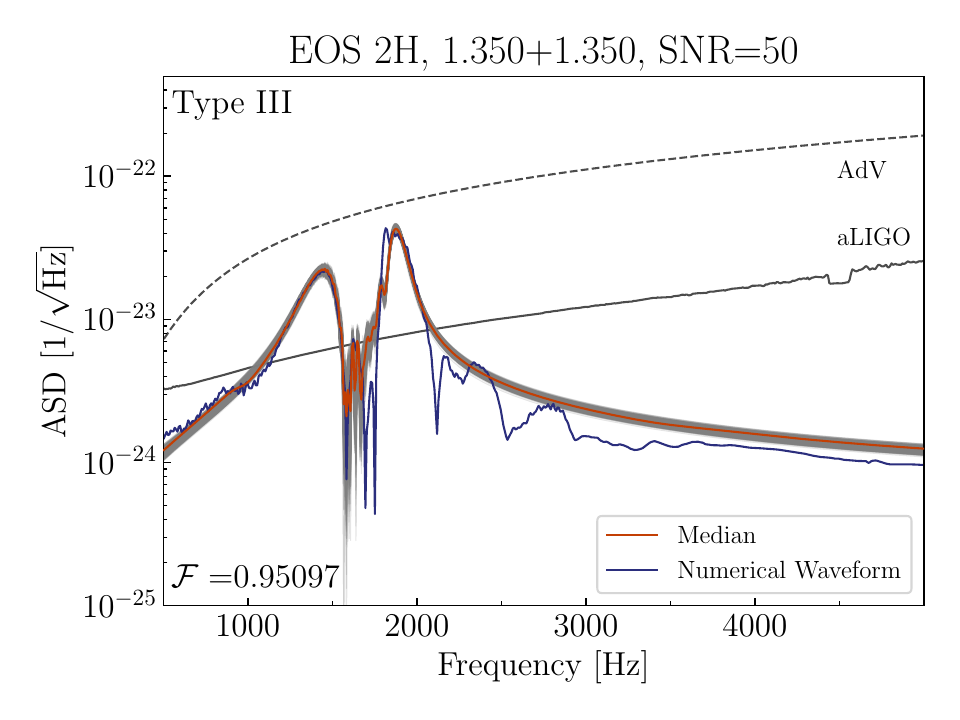} %
    \qquad
    \includegraphics[width=.50\textwidth]{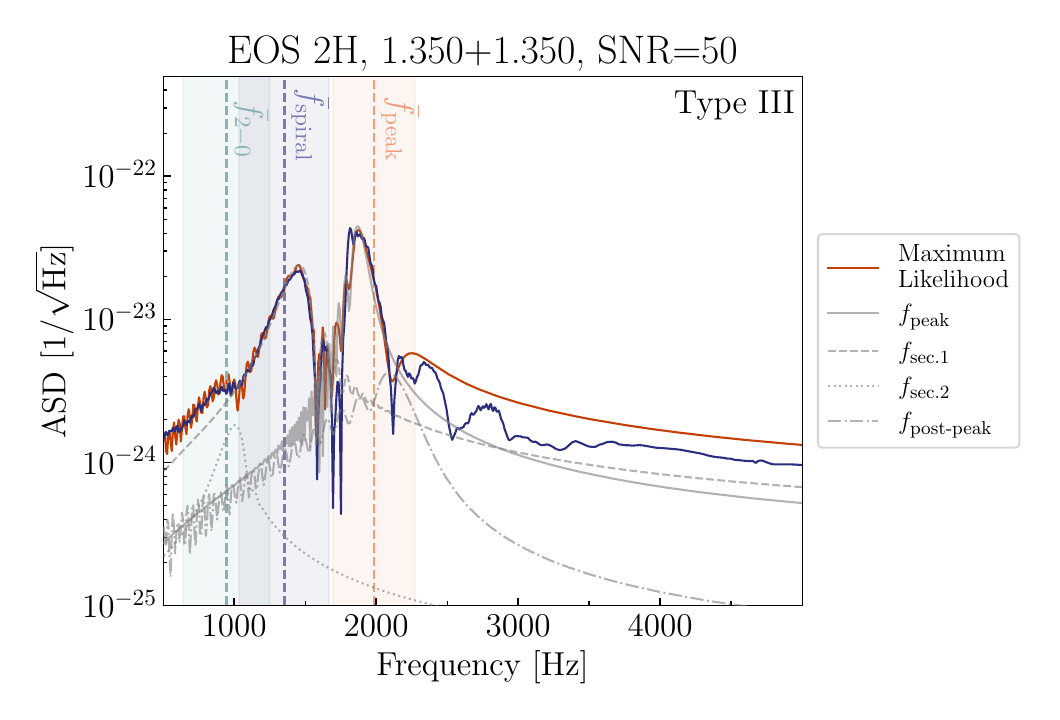} %
    \caption{Same as Fig. \ref{fig:SLy_1.5_1_medmax}, but for a 1.35+1.35 $M_\odot$ model with EOS 2H (Type III post-merger spectrum).}%
    \label{fig:2H-1.35_medmax}%
\end{figure*}

\subsubsection{Type II (EOS BHBlp, 1.3+1.3 $M_\odot$)} \label{sec:typeII}

Our third example is a 1.3+1.3 $M_\odot$ merger using the BHBlp EOS, which is well inside the Type II region in Fig.~\ref{fig:classification} and, therefore, both secondary peaks $f_{\rm spiral}$ and $f_{\rm 2-0}$ are important. 
In the left panel of Fig.~\ref{fig:BHBlp-1.3_medmax}, the 1000 randomly sampled waveforms (gray lines), constructed using the posterior distribution of the inferred parameters of our analytic model of Eq. (\ref{Eq:anmodel}) cluster tightly around their median (red line), in good agreement with the ASD of the numerical waveform (blue line), with larger variations only at low frequencies. The fitting factor between the median reconstructed waveform and the numerical waveform is ${\cal F} = 0.96336$. 

In the right panel of Fig.~\ref{fig:BHBlp-1.3_medmax}, all three main post-merger peaks, $f_{\rm peak}$, $f_{\rm spiral}$ and $f_{\rm 2-0}$ have frequencies close to the central frequency predicted by each corresponding empirical relation. The damped oscillator corresponding to $f_{\rm peak}$ has converged very well to the main $f_{\rm peak}$ contribution. The first secondary oscillator ($f_{\rm sec.1}$) describes well the $f_{\rm spiral}$ contribution and the other secondary oscillator ($f_{\rm sec.2}$) has converged to a frequency corresponding to the $f_{\rm 2-0}$ contribution. On the other hand, there is no strong feature at high frequencies and $f_{\rm post-peak}$ simply contributes a very broad peak (in practice the fourth oscillator could be left out of the analytic model in this case).

Overall, the analytic model of Eq. (\ref{Eq:anmodel}), when combined with the empirical relations and classification scheme, describe very well all the main features of the post-merger GW spectrum of type II models, and high fitting factors can be obtained.

\begin{figure*}[ht]
    \centering
    \includegraphics[width=.45\textwidth]{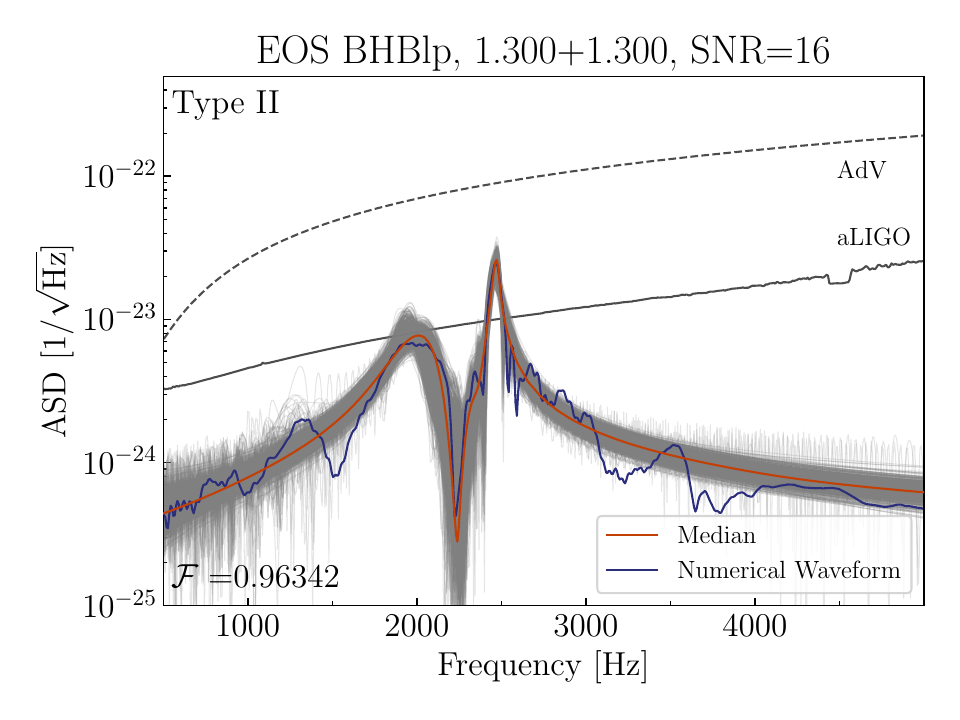} %
    \qquad
    \includegraphics[width=.50\textwidth]{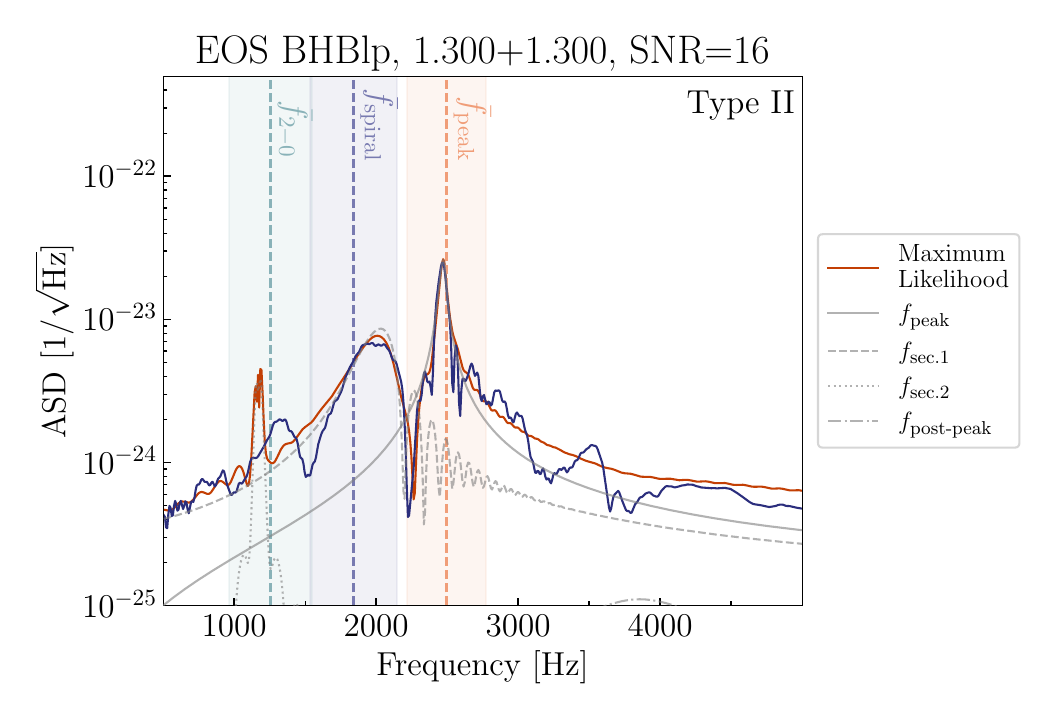} %
    \caption{Same as Fig. \ref{fig:BHBlp-1.3_medmax}, but for signal-to-noise ratio of 16.}%
    \label{fig:SNR16_medmax}%
\end{figure*}

\begin{figure*}[ht]
    \centering
    \includegraphics[width=.45\textwidth]{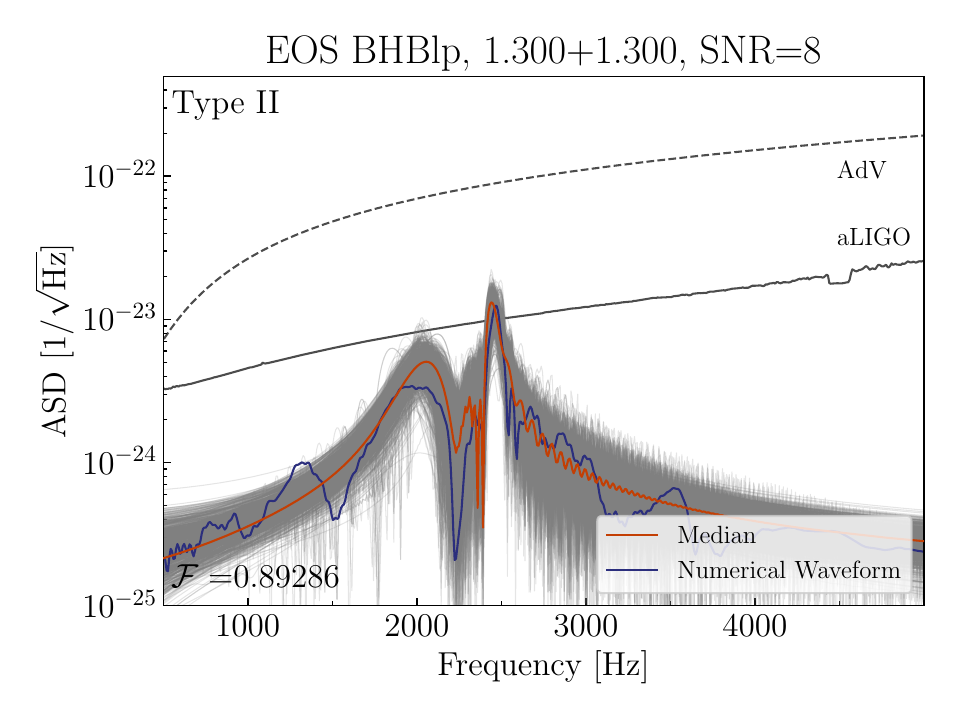} %
    \qquad
    \includegraphics[width=.50\textwidth]{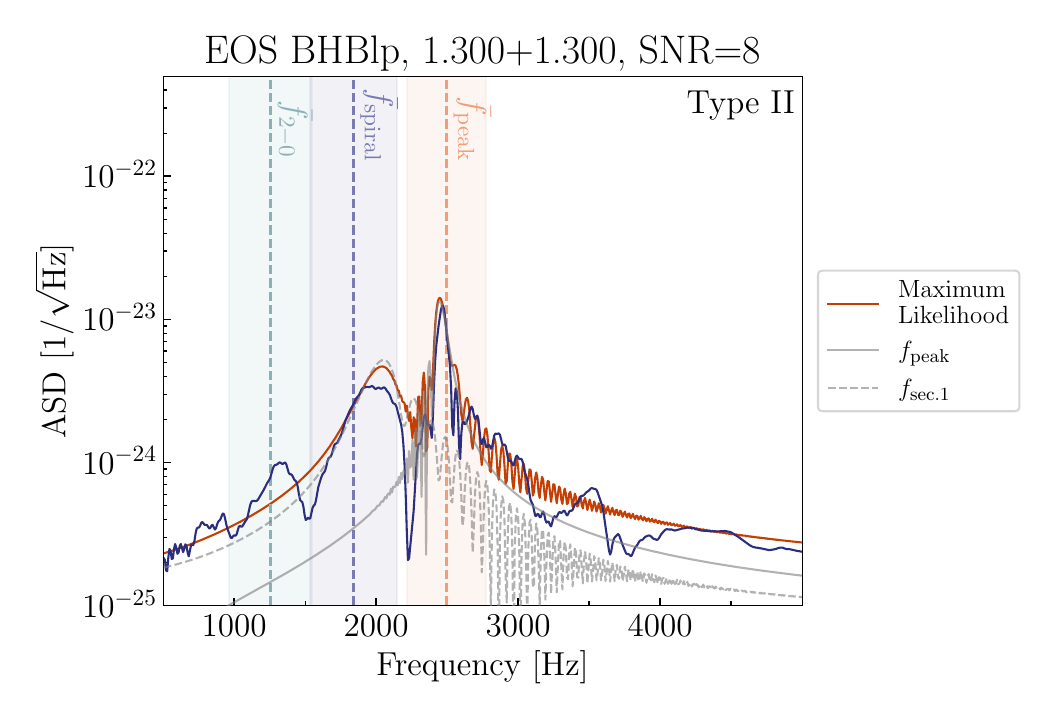} %
    \caption{Same as Fig. \ref{fig:BHBlp-1.3_medmax}, but for signal-to-noise ratio of 8.}%
    \label{fig:SNR8_medmax}%
\end{figure*}

\subsubsection{Type III (EOS 2H, 1.35+1.35 $M_\odot$)} \label{sec:typeIII}

Our last example is  a 1.35+1.35 $M_\odot$ merger with EOS 2H. This EOS is very stiff and the model lies deep inside the Type III region of Fig.~\ref{fig:classification}, i.e. $f_{\rm spiral}$ is the dominant secondary peak in the post-merger spectrum. Consequently, our scheme leads to Gaussian priors ${\cal N}(f_{\rm peak},\sigma^2)$, ${\cal N}(f_{\rm spiral},\sigma^2)$ centered on the predictions of the empirical relations (\ref{eq:peak}) and (\ref{eq:spiral}) for $f_{\rm fpeak}$ and $f_{\rm spiral}$, correspondingly, and uniform priors ${\cal U}(1, 5)$[kHz] for $f_{\rm 2-0}$. The fourth oscillator, $f_{\rm post-peak}$, in our analytic model (\ref{Eq:anmodel}) is initialized with uniform priors in the range $(f_{\rm min}, 5{\rm kHz})$, where $f_{\rm min}=f_{\rm peak}+0.3{\rm kHz}$.

In the left panel of Fig.~\ref{fig:2H-1.35_medmax}, we display the ASD of 1000 randomly sampled posterior waveforms (gray lines), which cluster very tightly around their median (red line) at all frequencies, in excellent agreement with the ASD of the numerical waveform (blue line). The fitting factor between the median reconstructed waveform and the numerical waveform is ${\cal F} = 0.95097$.

The right panel of Fig.~\ref{fig:2H-1.35_medmax} shows, in more detail, the individual contributions of the different damped oscillators included in the analytic model (\ref{Eq:anmodel}). 
We observe that the damped oscillator corresponding to $f_{\rm peak}$  describes the shape of $f_{\rm peak}$, which lies well inside the band predicted by the empirical relation (\ref{eq:peak}) with high accuracy. The same can be said for the damped oscillator corresponding to $f_{\rm spiral}$. Although the $f_{\rm 2-0}$ peak is barely distinguishable in this Type III spectrum, the corresponding damped oscillator still converges to the correct frequency and amplitude (notice that the maximum likelihood waveform is the {\it sum} of the individual contributions). At high frequencies, a strong $f_{\rm 2+0}$ does not exist in the Type III spectrum, but several smaller peaks can be observed. Because we gave a wide frequency range as prior to the fourth oscillator, $f_{\rm post-peak}$, it does not converge to a particular secondary peak at high frequencies, but it contributes to the reconstructed waveform over a wide frequency range.

Overall, the analytic model of Eq. (\ref{Eq:anmodel}), when combined with the empirical relations and classification scheme, describe very well the main features of the post-merger GW spectrum of type III models, and high fitting factors are obtained.

\subsection{${\rm SNR}=16$}

When the source is at a larger distance than considered in Sec. \ref{SNR50}, the signal to noise ratio drops. Here, we examine the impact of a lower SNR = 16 on the reconstruction of the post-merger spectrum for the 1.3+1.3 $M_\odot$ type II merger using the BHBlp EOS. We chose this model because both $f_{2-0}$ and $f_{\rm spiral}$ are important and could be reconstructed at ${\rm SNR}=50$ (although $f_{2-0}$ had greater uncertainty). 

In the left panel of Fig.~\ref{fig:SNR16_medmax}, the 1000 randomly sampled waveforms (gray lines) have a significantly wider distribution, compared to the corresponding ${\rm SNR}=50$ case in Sec. \ref{SNR50}. The median reconstructed waveform only informs about $f_{\rm peak}$ and $f_{\rm spiral}$, while the $f_{2-0}$ is no longer visible in the reconstruction. The fitting factor is still high, ${\cal F} = 0.96342$, due to the good reconstruction of the main peak.

The right panel of Fig.~\ref{fig:SNR16_medmax}, confirms that only two post-merger peaks, $f_{\rm peak}$ and $f_{\rm spiral}$ are reconstructed, with frequencies close to the central frequency predicted by each corresponding empirical relation. The damped oscillator corresponding to $f_{\rm peak}$ has converged very well to the main $f_{\rm peak}$ contribution. The first secondary oscillator ($f_{\rm sec.1}$) describes well the $f_{\rm spiral}$ contribution, but the other secondary oscillator ($f_{\rm sec.2}$) has converged to a frequency peak close to $f_{\rm 2-0}$, but with an incorrect amplitude and damping timescale. The fourth oscillator in the model $f_{\rm post-peak}$ does not converge. In practice, we find that for ${\rm SNR}=16$, a simpler analytic model employing only two oscillators, corresponding to $f_{\rm peak}$ and $f_{\rm spiral}$ would suffice to describe the post-merger spectrum for this merger.

% \begin{figure}
%     \centering
%     \includegraphics[width=.5\textwidth]{plots/SNR16/trace_plot.png}
%     \caption{Traceplot of BHBlp-M1.3 for SNR of 16.}
%     \label{fig:trace}
% \end{figure}

% \begin{figure*}
%     \centering
%     \makebox[\textwidth][c]{\includegraphics[width=1\textwidth]{plots/SNR16/joined_plot.pdf}}
%     \caption{Network response, reconstruction of injected waveforms and analysis of spectral diagram for the BHBlp-M1.3 waveform for SNR of 16.}
%     \label{fig:SNR16}
% \end{figure*}

\subsection{${\rm SNR}=8$}

For even larger source distances, when ${\rm SNR}=8$, the uncertainties in the reconstruction of the post-merger spectrum for the 1.3+1.3 $M_\odot$ type II merger using the BHBlp EOS increase significantly. Because of this, there is no point in using an analytic model with four damped oscillators. Instead, we attempt to reconstruct the post-merger waveform using only two damped oscillators, corresponding to $f_{\rm peak}$ and $f_{\rm spiral}$.

In the left panel of Fig.~\ref{fig:SNR8_medmax}, the 1000 randomly sampled waveforms (gray lines) have a much wider distribution than for the ${\rm SNR}=16$ case and the fitting factor has dropped to ${\cal F} = 0.89286$. However, the median reconstructed waveform still informs about both the $f_{\rm peak}$ and $f_{\rm spiral}$ peaks.

The right panel of Fig.~\ref{fig:SNR8_medmax}, confirms that both post-merger peaks, $f_{\rm peak}$ and $f_{\rm spiral}$ are reconstructed fairly accurately, with frequencies close to the central frequency predicted by each corresponding empirical relation. The damped oscillator corresponding to $f_{\rm peak}$ has converged very well to the main $f_{\rm peak}$ contribution and the secondary oscillator ($f_{\rm sec.1}$) describes well the $f_{\rm spiral}$ contribution.

 \section{Fitting Factors and Reconstruction}
\begin{figure}
    \centering
    \includegraphics[width=.5\textwidth]{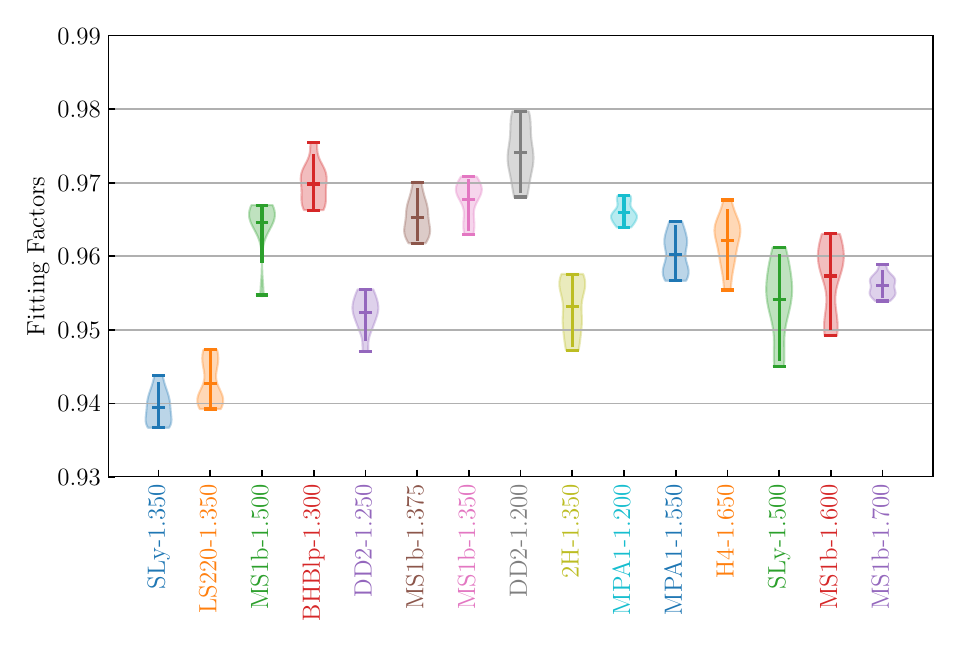}
    \caption{Fitting-factor distributions for each post-merger numerical-relativity waveform. 99.7\% confidence intervals are represented by the upper and lower horizontal bars. The median value appears as the horizontal bar in the centre. The 95\% confidence intervals are indicated by the thick vertical line. The range of the median fitting factors is 0.935 to 0.980.}
    \label{fig:FFs}
\end{figure}

We can evaluate the effectiveness of our methodology by computing the noise-weighted fitting factor of Eq. (\ref{eq:FF}) for each of the 15 cases we consider. Fig. \ref{fig:FFs} shows the distribution of the fitting factors in our sample of models. In each case, the upper and lower horizontal lines show the 99.7\% confidence intervals, based on 10 different noise realizations, with an intermediate line denoting the median value. In addition, a vertical line indicates the 95\% confidence interval. The median fitting factors are in the range of 0.935 to 0.980, similar to the results in \cite{Easter2020}.

\label{section:results}

\begin{figure}
    \centering
    \includegraphics[width=.5\textwidth]{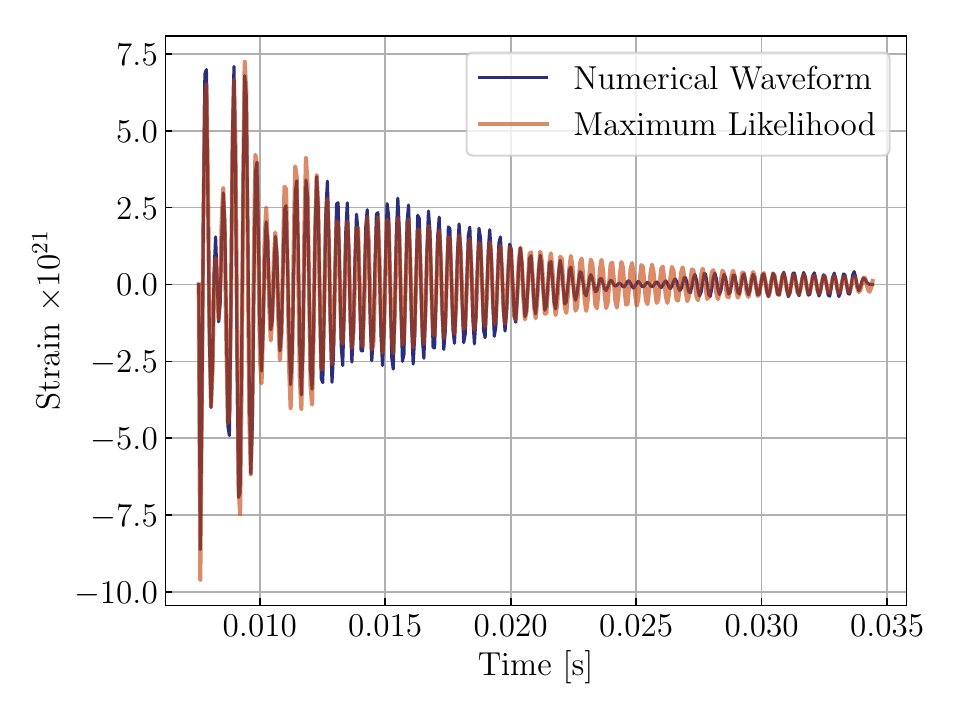}
    \caption{Scaled data alongside the maximum likelihood reconstruction time domain plot for the 1.3+1.3 BHBlp EOS.} 
    \label{fig:time_dom}
\end{figure}

A representative case of the reconstruction of the post-merger waveform using the maximum likelihood waveform is shown in Fig. \ref{fig:time_dom}, which is employed for the 1.3+1.3 BHBlp data. The reconstruction agrees well with the numerical waveform up to about 20ms after merger. The differences after that can be explained as follows: The reconstructed waveform is consistent with the assumption of the analytic model that the signal comprised several damped oscillators. On the other hand, the numerical waveform shows signs of the revival of the quadrupole frequency after about 25ms from merger. This behaviour is not built into the analytic model. The revival could be due to e.g. rotational instabilities developing in the remnant (see \cite{Sasli2023wpa} and references therein).

At a post-merger network SNR of 50, the posterior distributions show no multi-modal behaviour, and the typical $1\sigma$ error for extracting the $f_{\rm peak}$ frequency is $\sim 0.22\%$. For the dominant secondary peak $f_{\rm sec.1}$, the $1\sigma$ error is $\sim 1.4\%$ and for the weaker $f_{\rm sec.2}$ and $f_{\rm post-peak}$ frequencies it is $\sim 6.2\%$.
At lower post-merger network SNR values, we still typically find no multi-modal behavior in the posterior distributions.
The $1\sigma$ errors are  0.72\%, 2.6\%, 7.3\% and 18\% for $f_{\rm peak}$, $f_{\rm sec.1}$, $f_{\rm sec.2}$ and $f_{\rm post-peak}$ respectively, when SNR=16. For the weakest case of SNR=8, 
we find that  $f_{\rm peak}$ can still be extracted with $1\sigma$ of $2.4\%$, which increases to $8.5\%$ for the dominant secondary frequency.

 \section{Summary and Discussion}
 \label{sec:discussion}
 We propose an improved method for parameter estimation in the post-merger phase of binary neutron star mergers. The improvements come from the application of empirical relations for the three main post-merger frequencies to the choice of priors, as well as from the application of a classification of the different post-merger spectra. 
Our dataset consists of 15 waveforms of binary neutron star mergers that form differentially rotating remnants, 13 of which are from the \texttt{CoRe} catalog and two from \cite{soultanis2022}.
Most of our waveforms correspond to equal-mass cases, with the exception of two cases that have a mass ratio of 1.5.
The individual neutron star  radii in our dataset cover a range of $\sim11.5$km to $16$km and masses from $\sim1.2M_\odot$ to $1.7M_\odot$.
This range accommodates all three waveform types of the classification scheme introduced in \cite{Bauswein2015,VSB2020}, as can be seen in Fig. \ref{fig:classification}.
 
The analytic model we use to describe gravitational waves in the post-merger phase is an extension of the model analytical model in \cite{Easter2020}. Apart from the dominant $f_{\rm peak}$ frequency and the two subdominant frequencies $f_{\rm spiral}$, $f_{2-0}$ modes we also allow for the existence of a higher subdominant frequency. 

The main objective of our study is to develop a robust framework that systematically leads to converged posterior distributions of the main parameters involved in the analytic model, even when choosing different noise realizations. In addition, our objective was to resolve the appearance of multimodal posterior distributions, seen previously in some cases studied in \cite{Easter2020}.

Following \cite{VSB2020}, we derived additional empirical relations for post-merger frequencies, in terms of the tidal deformability ($\tilde\Lambda$) and the chirp mass ($\mathcal{M}_{\rm chirp}$), which are observables that can be determined from the inspiral phase. This allows us to set significantly smaller prior ranges for several parameters, which avoids convergence to wrong or multiple solutions. The classification scheme of 
\cite{Bauswein2015,VSB2020} is used to first determine the type of post-merger spectrum and then choose an approriate strategy for choosing the priors or leaving out certain components of the analytic model. Depending on the type of the post-merger spectrum, we use Gaussian priors for the dominant and main subdominant frequency peaks and uniform priors only for the weakest oscillators in the analytic model. In this way, the parameter estimation is optimized, taking into account the available knowledge from the inspiral phase and our current understanding of the post-merger GW spectrum.

Using different components in the analytic model, depending on type of the post-merger waveform, allowed us to identify the resolved peaks with particular physical processes in the merger remnant. This is important for testing the detailed physical assumptions that are made when simulating binary neutron star mergers.

Overall, we demonstrate that
 the analytic model of Eq. (\ref{Eq:anmodel}), when combined with the empirical relations, the choice of priors and the classification scheme, describes well the main features of the post-merger GW spectra of the three main types I, II and III and high fitting factors are obtained. In addition, we show 
 that our scheme  has sufficient flexibility to capture all essential features of the post-merger GW spectrum of Type Ib models, also reaching high fitting factors in this case.

Transitioning from \texttt{dynesty} to \texttt{pocoMC} as our primary sampler resulted in an acceleration of nearly tenfold in terms of walltime. The enhanced sampling performance offered by \texttt{pocoMC} arises from several contributing factors. Firstly, the efficiency of sampling is increased in the latent space of the normalizing flow due to the lack of strong correlations between the parameters of the posterior distribution. This eliminates the necessity of segregating the posterior into distinct sections - often accomplished via an ellipsoid mixture within the nested sampling context - and the subsequent need to sample each region individually. Additionally, the application of broad priors on certain parameters means that the prior volume substantially exceeds the posterior volume. This disparity significantly impairs the sampling performance of nested sampling methods, though the impact on PMC implementations such as \texttt{pocoMC} is less pronounced. Lastly, the inherent parallelizability of \texttt{pocoMC} enabled us to harness the power of multiple CPUs to further boost the sampling speed. This increased pace allowed for the exploration of a wider range of cases and a large number of noise realizations. We have thus demonstrated that \texttt{pocoMC} is an outstanding tool for parameter estimation of gravitational wave signals.

Here, we presented a robust and fast parameter estimation method of post-merger GW signals, based on a specific analytic model for the post-merger phase and on numerical waveforms that were obtained through numerical-relativity simulations where only the general-relativistic hydrodynamics was evolved. An alternative approach to analytic models is the use of machine-learning representations of the post-merger GW spectrum \cite{2022PhRvD.105l4021W,2024PhRvD.110f3008P,Soultanis2025}.  

A number of additional physical effects are expected to impact the waveform produced in the post-merger phase, e.g. turbulence modeling \cite{2024LRCA...10....1R}, magnetic fields \cite{2022PhRvD.105j4028S,2024PhRvD.110b4046B,2024arXiv241100939T,2024arXiv241100943B}, neutrino transport \cite{18Zappa2023,2022PhRvD.105j4028S,2024arXiv240909147P}, bulk viscosity \cite{2024ApJ...967L..14M}, dark matter imprints \cite{2024arXiv240805226S}, etc. These effects are expected to have an impact on the onset (or suppression) of late-time convective or rotational instabilities in the remnant \cite{DePietri2018,2020PhRvD.101f4052D,Sasli2023wpa,2023PhRvD.107j3053M}. Furthermore, deviations from general relativity are also expected to influence the structure of the post-merger remnant and affect the detected waveforms, see e.g. \cite{2013PhRvD..87h1506B,2014PhRvD..89h4005S,2018PhRvD..97f4016S,2021PhRvD.104j4036M,2022PhRvD.106j4055E,2022PhRvL.128i1103B,2023PhRvD.108b4058S,2023PhRvD.108f3033K, PhysRevD.87.081506,PhysRevD.89.044024,PhysRevD.89.084005,PhysRevD.97.064016,PhysRevD.104.104036,PhysRevD.106.104055,PhysRevD.108.064057,PhysRevLett.128.091103,PhysRevD.108.024058,PhysRevD.108.063033,2024PhRvD.110j4018L,2024arXiv241000137T}. In future studies, our current method could be tested for robustness against the impact of such effects.

%finito

\section{Acknowledgements}
We want to give special thanks Andreas Bauswein, Nikolaos Karnesis, Theodoros Soultanis and Nikolaos Chatzarakis, 
for insightful discussions. We are grateful to Sebastiano Bernuzzi for his comments on the manuscript.
Results presented in this work have been produced using the Aristotle University
of Thessaloniki (AUTh) High Performance Computing Infrastructure and Resources.
This material is based upon work supported by NSF's LIGO Laboratory which is a major facility fully funded by the National Science Foundation. We acknowledge support by Virgo, which is funded, through the European Gravitational Observatory (EGO), by the French Centre National de Recherche Scientifique (CNRS), the Italian Istituto Nazionale di Fisica Nucleare (INFN) and the Dutch Nikhef, with contributions by institutions from Belgium, Germany, Greece, Hungary, Ireland, Japan, Monaco, Poland, Portugal, Spain.

\nocite{*}

%Control: production of eprint (1) enabled
%

% \bibliography{bibliography.bib}% Produces the bibliography via BibTeX.

\end{document}